\def\ANON{0} 
\def\ICAARTmode{1} 
\def\DRAFT{0} 
\def\arXiv{1} 

\ifnum\ICAARTmode=1
  \documentclass[a4paper,twoside]{article} 
\else
  \documentclass[conference]{IEEEtran} 
\fi

\usepackage{epsfig}
\usepackage{subcaption}
\usepackage{calc}
\usepackage{amssymb}
\usepackage{amstext}
\usepackage{amsmath}
\usepackage{amsthm}
\usepackage{multicol}
\usepackage{pslatex}
\usepackage{natbib}
\usepackage[ruled,vlined]{algorithm2e}
\usepackage{balance}
\usepackage{flushend} 
\usepackage{booktabs} 

\usepackage{fancyhdr}

\ifnum\DRAFT=1
	\usepackage{draftwatermark} 
	\SetWatermarkText{
		\shortstack{
        			Draft \\[1em]  ICAART submission. \\[1em]  Do not distribute.
		}
	}
	\SetWatermarkLightness{0.8}
	\SetWatermarkScale{0.4}
\fi

\usepackage{SCITEPRESS}     

\ifnum\arXiv=1
\fi

\usepackage{xspace}
\usepackage{xcolor}
\definecolor{orange9}{HTML}{FFDD00}
\usepackage[color=orange9]{todonotes}

\begin{document}


\title{Fools Rush In: Competitive Effects of Reaction Time in Automated Trading}


\ifnum\ANON=1 

\author{\authorname{First Author\sup{1}\orcidAuthor{0000-0000-0000-0000} and Second Author\sup{1}\orcidAuthor{0000-0000-0000-0000}}
\affiliation{\sup{1}XXX, YYY, ZZZ}
\email{f\_author@xxx.yyy.zzz, s\_author@aaa.bbb.ccc}
}

\else 

\author{\authorname{Henry Hanifan\sup{1} and John Cartlidge\sup{1}\orcidAuthor{0000-0002-3143-6355}}
\affiliation{\sup{1}Department of Computer Science,University of Bristol, Merchant Venturers Building, Woodland Road, Bristol, UK}
\email{hh15092@my.bristol.ac.uk, john.cartlidge@bristol.ac.uk}
}

\fi 


\abstract{
We explore the competitive effects of reaction time of automated trading strategies in simulated financial markets containing a single exchange with public limit order book and continuous double auction matching. A large body of research conducted over several decades has been devoted to trading agent design and simulation, but the majority of this work focuses on pricing strategy and does not consider the time taken for these strategies to compute. In real-world financial markets, speed is known to heavily influence the design of automated trading algorithms, with the generally accepted wisdom that faster is better. 
Here, we introduce increasingly realistic models of trading speed and profile the computation times of a suite of eminent trading algorithms from the literature. Results demonstrate that: (a) trading performance is impacted by speed, but faster is not always better; (b) the Adaptive-Aggressive (AA) algorithm, until recently considered the most dominant trading strategy in the literature, is outperformed by the simplistic Shaver (SHVR) strategy---shave one tick off the current best bid or ask---when relative computation times are accurately simulated.  
}



\onecolumn \maketitle \normalsize \setcounter{footnote}{0} \vfill

%




\section{\uppercase{Introduction}}
\label{sec:introduction}
\noindent 
As trading in financial markets has become increasingly automated, the importance of speed is paramount. Competition between automated trading systems (ATS) looking to capitalise on fleeting opportunities ahead of rivals has resulted in a proliferation of high frequency trading (HFT) algorithms capable of executing many thousands of trades each second \citep{DuffinCartlidge18}.
The effects of ever-faster ATS (the so called {\em race to zero}) can be observed in the dynamics of modern financial markets: individual stocks now frequently exhibit ten percent price swings in less than one tenth of a second \citep{JohnsonEtal13}; flash crashes can cause whole markets to lose a trillion dollars in five minutes \citep{BaxterCartlidge13}; and when market-leading ATS's malfunction, the owners can be pushed to bankruptcy in under an hour \citep{BaxterCartlidge13}.

Yet, the literature on automated financial trading agents is largely bereft of considerations of computational speed, with the majority of work focusing on pricing strategies. 
One reason for this is largely historical, with trading agent experiments tending to follow Vernon Smith's seminal design that helped birth the field of experimental economics \citep{Smith62}. Using human participants, \cite{Smith62} set up a simple marketplace where traders took turns to quote prices and execute trades in an open outcry style marketplace. To enable strict comparisons with earlier work, later studies that introduced new automated trading algorithms (e.g., ZIC \citep{GodeSunder93}; ZIP \citep{Cliff97}; GD \citep{GD97}; and AA \citep{Vytelingum06}) tended to follow Smith's original design. Although subsequent works have gradually introduced a series of adaptations, such as the use of an order book \citep{DasEtal01}, real-time experiments that include human and agent participants \citep{DasEtal01,DeLucaCliff11,DeLuca11,CartlidgeCliff12}, and more realistic market dynamics such as continuous replenishment of assignments \citep{DeLuca11,CartlidgeCliff12} and continuously varying equilibria \citep{Cliff19,SnashallCliff20}, the time taken for agents to compute their strategy has remained significantly under-studied. 

Several works have demonstrated that altering the design of trading agent experiments can raise doubt over how well previously established results translate to real world markets. This had led several authors to call for more experimental {\em realism} (e.g., \cite{DeLuca11,CartlidgeCliff12,CartlidgeCliff18,Cliff19,SnashallCliff20}).
Here, we address that challenge by introducing minimal models of computation time (`thinking' time, or `reaction' time, of traders) into the standard experimental framework. Using an adaptation of the {\em Bristol Stock Exchange} \citep{Cliff18-BSE}, we explore the competitive effects of computation speed on a suite of reference trading algorithms available on the platform: GVWY, SHVR, ZIC, ZIP, and AA. 

Perhaps unsurprisingly, results demonstrate that the relative speeds of trading strategies {\em does} affect profitability. However, depending on the particular strategies competing in the market, being faster is {\em not necessarily better}. Further, we show that when relative reaction times are accurately modelled, AA -- long considered the dominant trading agent strategy in the literature -- is beaten in static symmetric markets by the simple non-adaptive trading strategy SHVR. Given other recent evidence that AA's dominance is sensitive to the mixture of competing strategies in the market \citep{Vach15,Cliff19,SnashallCliff20}, and the complexity of the market dynamics \citep{Cliff19,SnashallCliff20} this result adds further support that AA is non-dominant when markets are more realistically modelled. 

The rest of the paper is organised as follows. In Section~\ref{sec:background} we review related work and introduce key economic concepts and technical details of the trading agents used in these experiments.
In Section~\ref{sec:method} we introduce three methods for modelling trader speed: (i) {\em fixed ordering}, (ii) {\em tournament ranking}, and (iii) {\em speed proportional selection}. Section~\ref{sec:results} presents results from a series of experiments with homogeneous markets (containing one trader type) and heterogeneous pair-wise balanced markets (containing two trader types).
In Section~\ref{sec:discussion} we discuss the implications of results and describe avenues for future work. Finally, Section~\ref{sec:conclusion} presents the conclusion that reaction speeds matter, and therefore the research community will benefit from future focus in this area. 

\section{\uppercase{Background}}
\label{sec:background}
\noindent 

\subsection{Trading Agent Experiments}
\label{sec:background-agents}
\noindent 
In the 1960s, Vernon Smith conducted a series of trading experiments with small groups of untrained human participants (i.e., students) to investigate competitive market behaviours \citep{Smith62}. He was able to demonstrate that these simple simulations of financial markets produced surprisingly efficient equilibration behaviours, with trade prices quickly tending to the theoretical equilibrium value predicted by the underlying market supply and demand. Intriguingly, three decades later, \cite{GodeSunder93}  were able to reproduce similar results, but this time using `zero intelligence' (ZI) algorithmic traders that generate random quote prices. Despite their simplicity, markets of ZIC traders (the letter C indicating traders are {\em constrained} to not make a loss) were shown to exhibit equilibration behaviours similar to that of humans, suggesting that intelligence is not necessary for competitive markets to behave efficiently: the market mechanism (the rules of the continuous double auction) performs much of the work.

\begin{figure}[tb]
  \centering
   \includegraphics[width=0.9\linewidth]{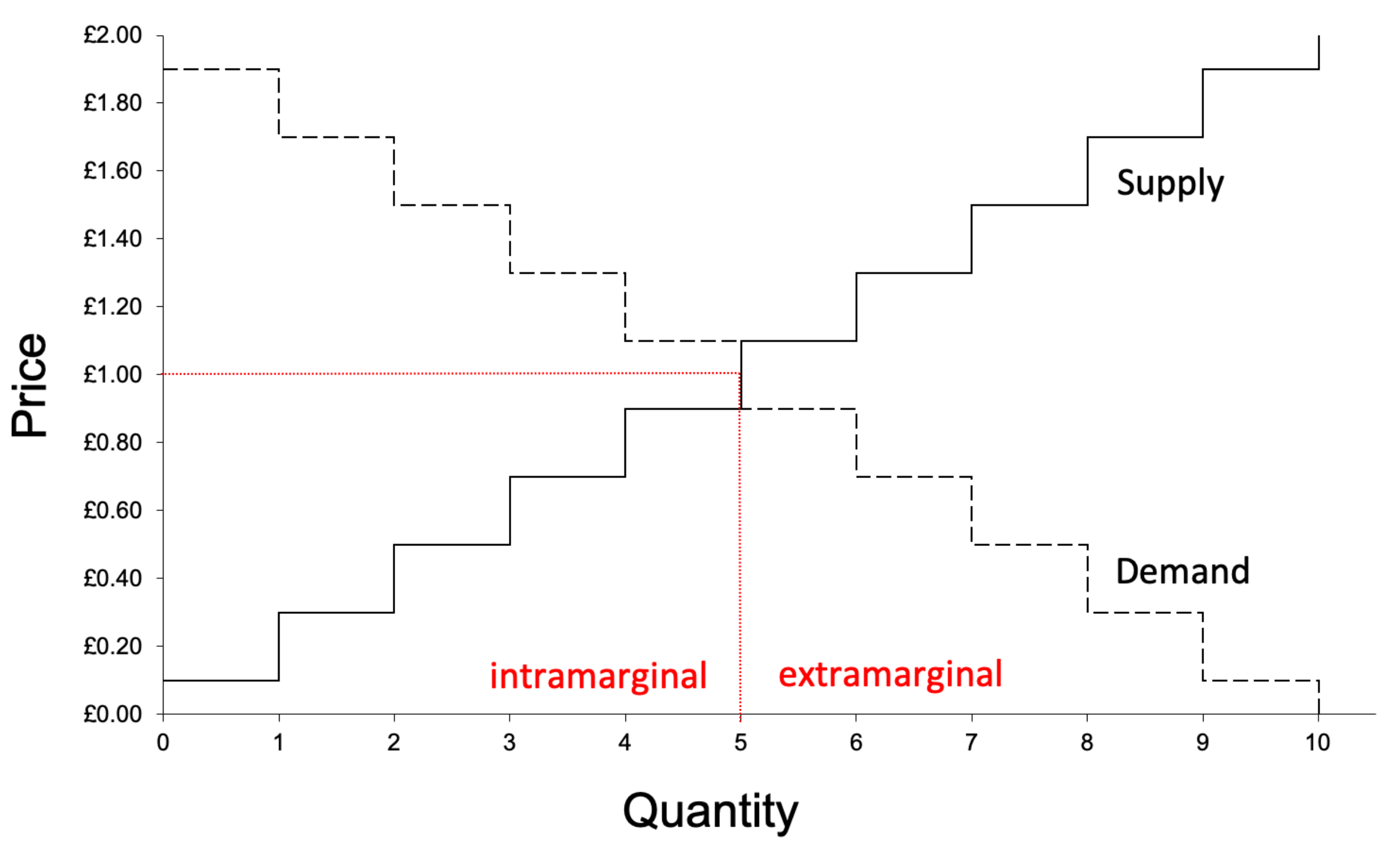}
  \caption{Symmetric supply and demand schedules, showing a market with $n=10$ buyers (demand) and $n=10$ sellers (supply).
  Limit prices for buyers (maximum price to buy) and sellers (minimum price to sell) are evenly distributed over the interval $[\pounds0.10, \pounds1.90]$, giving theoretical equilibrium price $P_0=\pounds1.00\pm0.10$ and expected quantity transacted $Q_0=n/2=5$. {\em Intra-marginal} traders, to the left of $Q_0$, expect to transact. {\em Extra-marginal} traders do not.}
  \label{fig:DS}
\end{figure}

However, Gode and Sunder's (1993) result was later shown to only hold when market demand and supply are symmetric (i.e., when the magnitude of the gradient---the {\em price elasticity}---of supply and demand schedules are similar, such as the example shown in Figure~\ref{fig:DS}). For asymmetric markets, such as when the supply curve is horizontal, `zero' intelligence is not enough to provide human-like levels of market efficiency \citep{Cliff97}. To account for this, Cliff introduced a new minimally-intelligent trading algorithm, which he named {\em Zero Intelligence Plus} (ZIP). ZIP maintains an internal profit margin, $\mu$, which is increased or decreased by traversing a decision tree that considers the most recent quote price, the direction of the quote (buy or sell) and whether it resulted in a trade. Margin, $\mu$, is then adjusted with magnitude proportional to a learning rate parameter, similar to that used in Widrow-Hoff or in back-propagation learning. \cite{Cliff97} successfully demonstrated that markets containing only ZIP traders will exhibit human-like behaviours in all of Smith's original experimental market configurations, both symmetric and asymmetric. 

\begin{figure}[tb]
  \centering
   \includegraphics[width=0.75\linewidth]{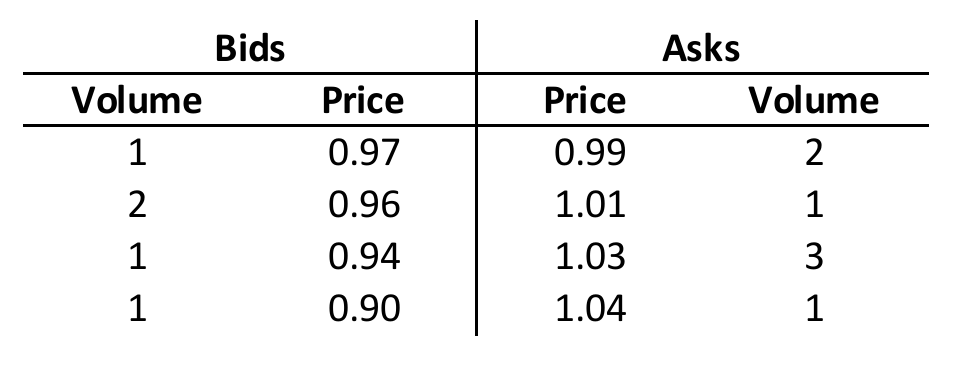}
  \caption{A Limit Order Book (LOB), presenting the current market state. Bids (orders to buy) are presented on the left hand side, ordered by price {\em descending}. Asks (orders to sell) are presented on the right hand side, ordered by price {\em ascending}.  Volume indicates the quantity available at each price. The top line presents the Best Bid ($BB=0.97$) and Best Ask ($BA=0.99$) prices in the market, and the difference between these prices is called the $spread=BA-BB=0.02$. The {\em midprice} of the book is $(BB+BA)/2=0.98$;
the {\em microprice} is volume weighted midprice, calculated as: $(2/3) 0.97+(1/3) 0.99=0.977$. Orders can be submitted at any price, subject to a minimum resolution, or tick size ($tick=0.01$). Aggressive orders that {\em cross the spread} (i.e., an ask with price $p_a\leq0.97$, or a bid with price $p_b\geq0.99$) will immediately execute at the price presented in the LOB (i.e., the ask will transact at price $BB=0.97$; the bid will transact at price $BA=0.99$). Passive orders that do not cross the spread will rest in the LOB, with position determined by price.}
  \label{fig:LOB}
\end{figure}

Other intelligent trading agents have been developed to maximise profits in experimental markets that follow Smith's framework. Most notably, these include: GD, named after its inventors, \cite{GD97}; and {\em Adaptive-Aggressive} (AA), developed by \cite{Vytelingum06}. GD selects a quote price by maximising a `belief' function of the likely profit for each possible quote, formed using historical quotes and transaction prices in the market. 
Over time, the original GD algorithm has been successively refined: first by \cite{DasEtal01} and \cite{TesauroDas01} (named {\em Modified GD}, or MGD) to enable trading using an order book (see example in Figure~\ref{fig:LOB}), and to reduce belief function volatility; and then by \cite{TesauroBredin02}, who used dynamic programming to optimise cumulative long-term discounted profitability rather than immediate profit ({\em GD eXtended}, or GDX).  
In contrast, AA incorporates a combination of short-term and long-term learning to update an internal profit margin, $\mu$. In the short-term, $\mu$ is updated using rules similar to ZIP. Over the long-term, AA calculates a moving average of historical transaction prices to estimate the market equilibrium value, $P_0$, and current price volatility calculated as root mean square deviation of transaction prices around the estimate $P_0$. If the AA trader estimates that it is extra-marginal (and will therefore find it difficult to trade profitably: see Figure~\ref{fig:DS}) it trades more aggressively (by reducing $\mu$), if it is intra-marginal (and will therefore find it easier to profit) it trades more passively (by increasing $\mu$).  

For a summary of trading strategies, see Table~\ref{tab:traders}.

\begin{table*}[tb]
\vspace{3mm}
\caption{Summary of trading agent strategies when acting as a buyer. When selling, prices are moved in the opposite direction. $Q$ is new quote price, $L$ is limit price, $T$ is tick size, $BB$ is best bid on the LOB. Traders cannot make a loss, i.e., $Q\leq L$.}\label{tab:agents} \centering
\small
{\renewcommand{\arraystretch}{1.15}
\begin{tabular}{ccp{0.67\textwidth}}
  \toprule
  Agent & Name & Method to Determine New Quote Price, $Q$ \\
  \midrule
  GVWY & Giveaway & $Q=L$. Always post quote at price equal to limit price.\\
  SHVR & Shaver & $Q=min(BB+T,L)$. Post quote one tick inside current best bid. \\
  ZIC & \begin{minipage}[t]{0.15\textwidth}\centering Zero Intelligence Constrained\end{minipage} & $Q=q\in U[0.01,L]$. Quote randomly from Uniform distribution bounded by system minimum value (one tick, $T=0.01$) and limit price, $L$. \\
  ZIP &  \begin{minipage}[t]{0.15\textwidth}\centering Zero Intelligence Plus\end{minipage} & $Q=L(1-\mu)$, where  $0\leq\mu<1$ is an internal profit margin. 
  When a trade occurs, if $Q$ is greater than trade price, then decrease $\mu$ (i.e., raise price), otherwise increase $\mu$. If new best bid on LOB has price $BB>Q$, decrease $\mu$ (i.e., raise price). \\
   AA & \begin{minipage}[t]{0.15\textwidth}\centering Adaptive Aggressive\end{minipage} & $Q=L(1-\mu)$, where  $0\leq\mu<1$ is an internal margin. Estimate market equilibrium, $P_0$, to determine whether $L$ is intra-marginal ($L\geq P_0$) or extra-marginal ($L<P_0$). If extra-marginal, increase aggressiveness (decrease $\mu$); else increase $\mu$. \\
   GDX & \begin{minipage}[t]{0.15\textwidth}\centering GD eXtended \\(not used in this study)\end{minipage}  & $Q$ selected by dynamic programming to maximise cumulative long-term discounted profitability of a `belief' function that calculates likely outcome of each price, $q$, based on the success of previous quotes and transaction prices in the market. \\
  \bottomrule
\end{tabular}
}
\label{tab:traders}
\end{table*}

\subsection{The Battle for Trading Dominance}\label{sec:background-dominance}
\noindent
For the last two decades, a research theme has emerged: to develop the best trading agent that can successfully beat human participants and other trading agents in Smith-style experiments (see \cite{SnashallCliff20} for a detailed historical account). It was first demonstrated that trading agents, specifically ZIP and MGD, outperform humans when directly competing in human-agent markets \citep{DasEtal01}: {\em ``\ldots the successful demonstration of machine superiority in the CDA and other
common auctions could have a much more direct and powerful financial impact---one that
might be measured in billions of dollars annually''}.
This announcement quickly generated global media coverage and significant industry interest.
Shortly afterwards, \cite{TesauroBredin02} suggested that GDX {\em ``may offer the best performance of any published CDA bidding strategy''}. Subsequently, after it's introduction in 2006 \citep{Vytelingum06}, AA was shown to dominate ZIP and GDX \citep{Vytelingum08} and also humans \citep{DeLucaCliff11}: {\em ``we therefore claim that AA may offer the best performance of any published strategy''}. And so, for several years, AA held the undisputed algo-trading crown.

However, more recently, doubt about the dominance of AA has emerged. In particular, for markets containing AA, GDX, and ZIP strategies, the mixture (i.e., the {\em proportion}) of strategies in the market has been shown to affect AA performance. In particular, AA only dominates when there is a significant proportion of other AA agents in the market; in other cases, it is regularly beaten by GDX and ZIP \citep{Vach15}. This finding was supported by \cite{Cliff19}, through exhaustive testing of markets containing mixtures of MAA (a slightly {\em modified} version of AA which utilises {\em microprice} of the orderbook; refer to Figure~\ref{fig:LOB}), ZIC, ZIP, and SHVR (a simple non-adaptive strategy that quotes prices one tick inside the current best price on the order book). Further,  \cite{Cliff19} found that introducing more realistic market dynamics---continuous replenishment of assignments, rather than periodic replenishments at regular intervals; and also a continuously moving equilibrium, $P_0$, which was set to follow real world historical trade price data---MAA did {\em not} dominate, and when considering {\em profitability}, MAA was significantly outperformed by ZIP and SHVR. A related study by \cite{SnashallCliff20} also showed that GDX dominates MAA, ZIP, ZIC, and ASAD ({\em Assignment-Adaptive}, developed by \cite{Stotter13}) in these more realistically complex markets (SHVR was not tested in the latter study). 

\subsection{Latency and Reaction Speed}\label{sec:background-speed}
\noindent
Throughout the previous works, the primary motivation has been focused on pricing strategies for trading efficiency (i.e., profit maximisation and market equilibration behaviours). However, if we are to better understand the behaviour of these algorithms in more realistic environments, it is important to consider {\em latency}, a key real-world factor that is missing in most of these studies. In real-world financial markets, {\em communication latency} (the differential delays in which traders can access trading information and initiate trades with an exchange), and {\em trading latency} (or {\em reaction time}: the time it takes for a human or algorithmic trader to react to new information) are major determinants of trading behaviours and market dynamics \citep{DuffinCartlidge18,SnashallCliff20}. In real markets, the proliferation and profitability of high frequency trading (HFT) evidences the efficacy of harnessing reduced latency, enabling traders to capitalise on fleeting opportunities ahead of competitors. 

Several studies have conducted human-agent and agent-agent trading experiments using real-time asynchronous trading platforms. For their seminal demonstration of agents outperforming human traders, \cite{DasEtal01} used a hybrid platform consisting of two of IBM's proprietary systems: GEM, a distributed experimental economics platform; and Magenta, an agent environment. Although real-time asynchronous, trading agents were constrained to operate on a sleep-wake cycle of $\bar{s}$ seconds, with {\em fast} agents having mean sleep time $\bar{s}=1$, and {\em slow} agents having mean sleep time $\bar{s}=5$. A random jitter was introduced for each sleep $s$ such that: $s\in[0.75\bar{s},1.25\bar{s}$]. Fast agents were set to wake on all new orders and trades, slow traders were set to wake only on trades. Therefore, although this real-time system enabled asynchronous actions, algorithmic traders were artificially slowed to have reaction times comparable with human traders. 

Following \cite{DasEtal01}, other real-time human-agent experiments have invoked a similar sleep-wake cycle. Using the {\em Open Exchange} (OpEx) platform,\footnote{OpEx: The Open Exchange. Available at: https://sourceforge.net/projects/open-exchange} 
\cite{DeLuca11} demonstrated AA, GDX, and ZIP outperform humans when agents have sleep-wake cycle $\bar s=1$; agent-agent experiments, demonstrating AA dominance, were performed using a discrete event model (such that reaction times were ignored). OpEx has also been used for further human-agent experiments, for example, to demonstrate that: aggressive ({\em spread-jumping}) agents that are faster (i.e., those with lower $\bar{s}$ values) can perform less well against humans \citep{DeLuca11}; faster trading agents can reduce the efficiencies of human traders in the market \citep{Cartlidge12}; and agents with reaction speeds much quicker than humans can lead to endogenous fragmentation within a single market, such that fast (slow)
traders are more likely to execute with fast (slow) traders (\cite{CartlidgeCliff12}; a result that has analogies with the {\em robot phase transition} demonstrated in real-world markets \citep{JohnsonEtal13}). 

Agent-only real-time asynchronous experiments have also been conducted using the {\em Exchange Portal} (ExPo) platform.\footnote{ExPo: The Exchange Portal. Available at: https://sourceforge.net/projects/exchangeportal} 
\cite{Stotter13,Stotter14} used ExPo to introduce a new Assignment-Adaptive (ASAD) trading agent. They demonstrated that in ASAD:ZIP markets (with sleep-wake cycle, $\bar{s}=4$), signals produced by the trading behaviour of ASAD are beneficially utilised by ZIP traders, to the detriment of ASAD themselves. 

Communications latency has been considered in other works, e.g.: \cite{DuffinCartlidge18} model latency arbitrage in fragmented markets; and \cite{MilesCliff19} study the effects of latency in simulated markets distributed globally in the cloud.  Trading speed has also been incorporated into strategies, e.g.,: \cite{gjerstad03} used a {\em pace} parameter to control the arrival rate of GD (renamed Heuristic Belief Learning, or HBL) traders, and to alter quote price as a function of elapsed time; and \cite{McGroarty19} introduced an agent model of financial markets with agents that operate on different timescales to simulate common strategies and behaviours, such as market makers, fundamental traders, high frequency momentum and mean-reversion traders, and noise traders. 

These works are representative of the literature relevant to reaction time in trading algorithms. In all, we see that computation times are either skewed by enforced sleep, directly encoded, or drawn from a probability distribution. As far as the authors are aware, there is no attempt to systematically understand the effects of reaction time using accurate computation times of individual strategies. Here, we attempt to address this gap.

\section{\uppercase{Methodology}}
\label{sec:method}
\noindent 

\subsection{BSE: The Bristol Stock Exchange}
\label{sec:bse}
The {\em Bristol Stock Exchange} (BSE) is a minimal, discrete-time simulation of a centralised financial market, containing a single exchange with Limit Order Book (LOB), and reference implementations of several leading trading strategies from the literature, including five trading strategies that we consider in this paper: GVWY, SHVR, ZIC, ZIP, and AA. For a summary description of each trader, see Table~\ref{tab:traders}. For full details on BSE, refer to \cite{Cliff18-BSE}.\footnote{BSE: The Bristol Stock Exchange. Available at: https://github.com/davecliff/BristolStockExchange. In this paper, we use BSE version 22/07/18, commit hash: c0b6a1080b6f0804a373dbe430e34d062dc23ffb.}

\subsubsection{Random Order Selection (BSE Default)}
\label{sec:method-random}
Each time step, BSE ensures all traders act exactly once by selecting traders at random, without replacement. As such, there is no concept of relative reaction times of traders. Over the course of a simulation experiment, a fast trader will have the same number of opportunities to act as a slow trader. Due to the randomised sequence of actions, the slower trader will be able to act first when presented with `lucky' opportunities as often as the fast trader. This is unrealistic. In the following section, we introduce three simple models of reaction time to the BSE framework.

\subsection{Modelling Reaction Time}
\label{sec:models}

\subsubsection{Fixed Order Selection}
\label{sec:method-order}
The simplest model of reaction time that we implement is the ordering model. Here, traders are selected to update and act in a fixed order, alternating each time step between buyers first and sellers first to ensure bias is not introduced. For each buyer $b_i\in B$ and seller $s_i\in S$ (where $|B|=|S|=n$), we introduce two orderings: 

\begin{equation}
Order_A = s_1, b_1, s_2, b_2, \ldots, s_n, b_n
\end{equation}
\begin{equation}
Order_B = b_1, s_1, b_2, s_2,\ldots, b_n, s_n
\end{equation}

In the first time step, one of the orderings is selected at random (e.g, $Order_A$); in the second time step, the ordering is switched (i.e., $Order_B$); the third time step returns to the first ordering ($Order_A$); etc. In this way, traders $s_1$ and $b_1$ are always selected to act first each time step; $s_n$ and $b_n$ are always selected to act last. We can consider this as traders $s_1$ and $b_1$ as acting much quicker (having a smaller computation time, or reaction time) than traders $s_n$ and $b_n$.  

\subsubsection{Tournament Ranking Selection}
\label{sec:method-rank}
The second model of reaction time assigns a speed ranking to each trader. Each time step, we perform the routine presented in Algorithm~\ref{alg:rank}. We begin by creating a pool containing all traders, and then select two traders at random from the pool and compare rank. The trader with the lowest rank is selected to act and then removed from the pool. This simulates two traders racing to act on new market information. We repeat these steps until all traders have acted and the pool is empty. Similar to the previous random order (Section~\ref{sec:method-random}) and fixed order (Section~\ref{sec:method-order}) models, this process ensures that all traders act exactly once each time step.

\begin{algorithm}[t]
\small
\SetAlgoLined
\SetKwInOut{Input}{input}\SetKwInOut{Output}{output}
\KwResult{All traders have acted exactly once}
\Input{$P$ is the set of all traders, size $2n$}
 \While{(size(P) $>$ 1)}{
  Randomly select trader $A$ from $P$\;
  Randomly select trader $B$ from $P$\;  
  \eIf{(rank(B) $>$ rank(A))}{
   Trader $A$ to act\;
   Remove $A$ from set $P$\;
   }{
   Trader $B$ to act\;
   Remove $B$ from set $P$\;
  }
 }
 Select remaining trader in $P$ to act\;
 \caption{Tournament Ranking}
 \label{alg:rank}
\end{algorithm}

\subsubsection{Speed Proportional Selection}
\label{sec:method-time}
The final model of reaction time is the most realistic. Every trader is initially assigned a reaction time. Each time step, traders are selected to act in proportion to their relative speeds, such that trader $A$, with a reaction time of $R^A=1$ would act twice as often as trader $B$ with a reaction time of $R^B=2$. To achieve this, each time step we select traders from a biased pool containing multiple references to each trader, such that the number of references is inversely proportional to each trader's relative reaction time. For example, if $R^A=1$ and $R^B=2$, we generate a biased pool, $P=\{A,A,B\}$, containing two references to trader $A$ and one reference to trader $B$. Each time step, traders are randomly selected to act, without replacement, until the pool is empty. 
We use notation $R^A_B=1/2$ to indicate A's reaction time is half B's reaction time; similarly $R^B_A=2$ indicates B's reaction time is twice as long as A. 

This model provides two advantages over the previous models. First, it enables faster agents to act multiple times before a slower agent can act (for $R^B_A=2$, each time step $A$ acts twice, while  $B$ acts only once). Second, it allows slower traders to occasionally get `lucky' by being selected first. Therefore, of the simple models presented here, this process most accurately simulates the real world.

\subsection{Experimental Configuration}
\label{sec:config}
For all experiments, markets contained an equal number of $n$ buyers and $n$ sellers, with assignment limit prices distributed evenly across the interval $[\pounds0.10,\pounds1.90]$, as shown in Figure~\ref{fig:DS}. Market sessions lasted 330 time steps, with assignments replenished periodically every 30 time steps.  Each simulation configuration was repeated 100 times. Error bars on graphs indicate 95\% confidence intervals. Statistical significance is calculated using Student's t-test. This simple, static, symmetric market is deliberately chosen to enable comparisons with the literature. Reaction time, using the procedures defined in Section~\ref{sec:models}, is the only independent variable that we manipulate, with the BSE default setting (all traders have equal reaction times) directly equivalent to previous studies.

\begin{figure}[tb]
  \centering 
   \includegraphics[width=0.7\linewidth]{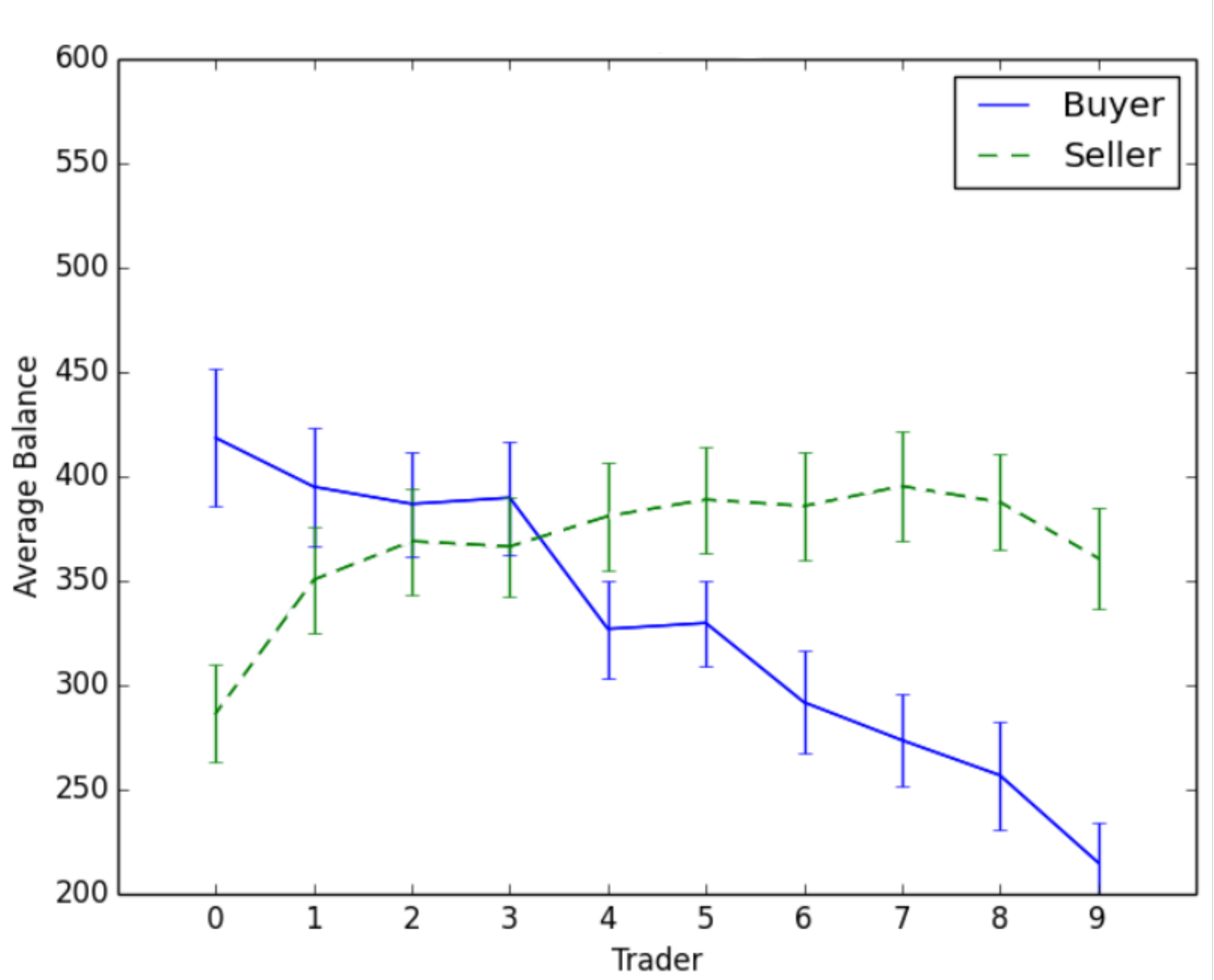}
  \caption{Fixed order selection for homogeneous ZIP markets.
  Buyers (solid blue line) perform better when selected earlier, while sellers
  (dashed green line) perform worse.} 
  \label{fig:homo-order-ZIP}
\end{figure}

\begin{figure}[tb]
  \centering 
  \includegraphics[width=0.7\linewidth]{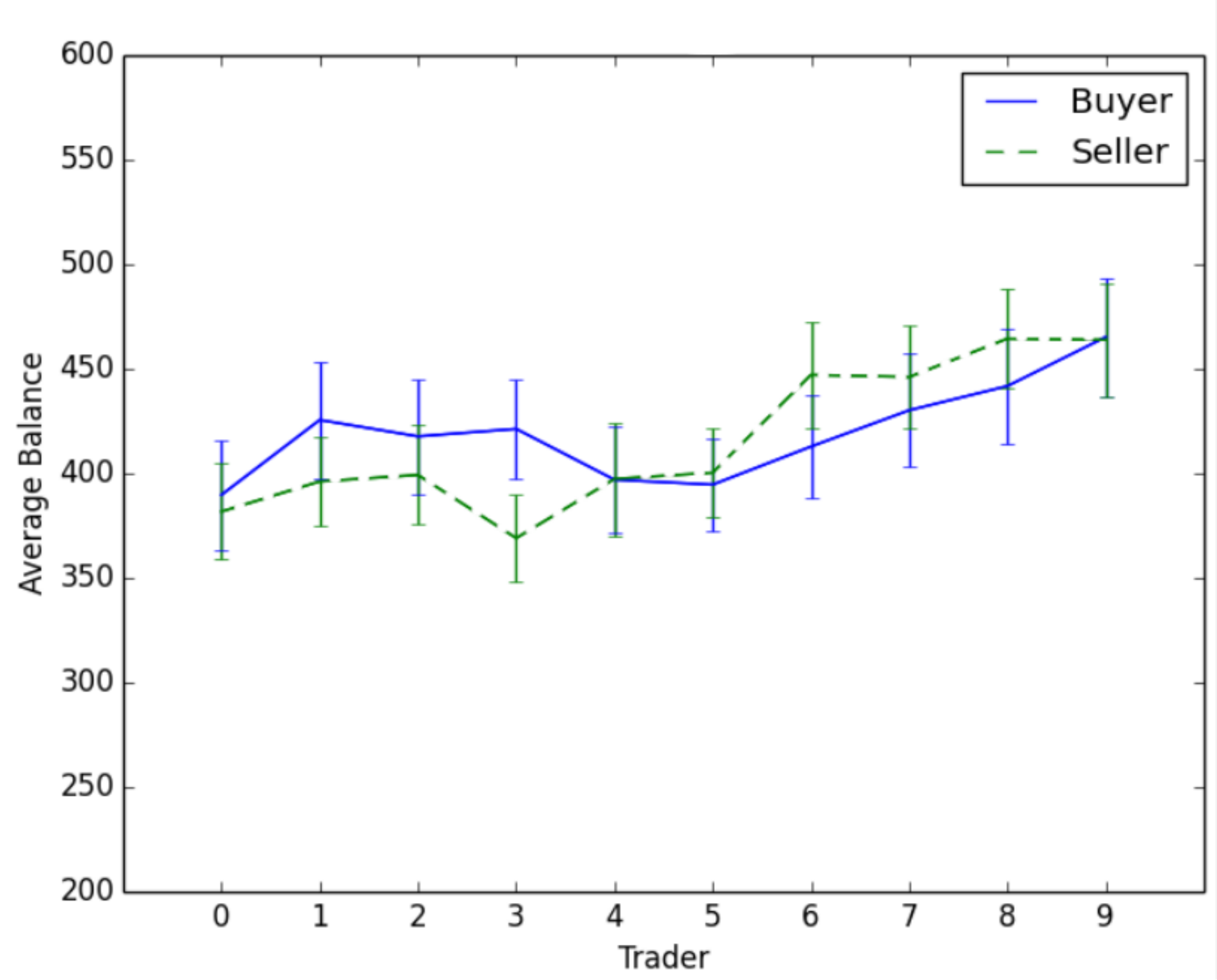}
  \caption{Fixed order selection for homogeneous AA markets.
  All traders tend to perform better when selected later.} 
  \label{fig:homo-order-AA}
\end{figure}

\section{\uppercase{Results}}
\label{sec:results}
\noindent 

\ifnum\ANON=0 
{
\noindent
We present results of experiments originally performed for an MSc project. 
For further details, see \cite{Hanifan19}.
}
\else
{
\noindent
For further details see [{\em citation removed for blind review}].
}
\fi 

\ifnum\arXiv=0
    \begin{figure*}[tb]
     \centering 
      \subcaptionbox{AA Profits}[0.32\linewidth][c]{%
        \includegraphics[width=\linewidth]{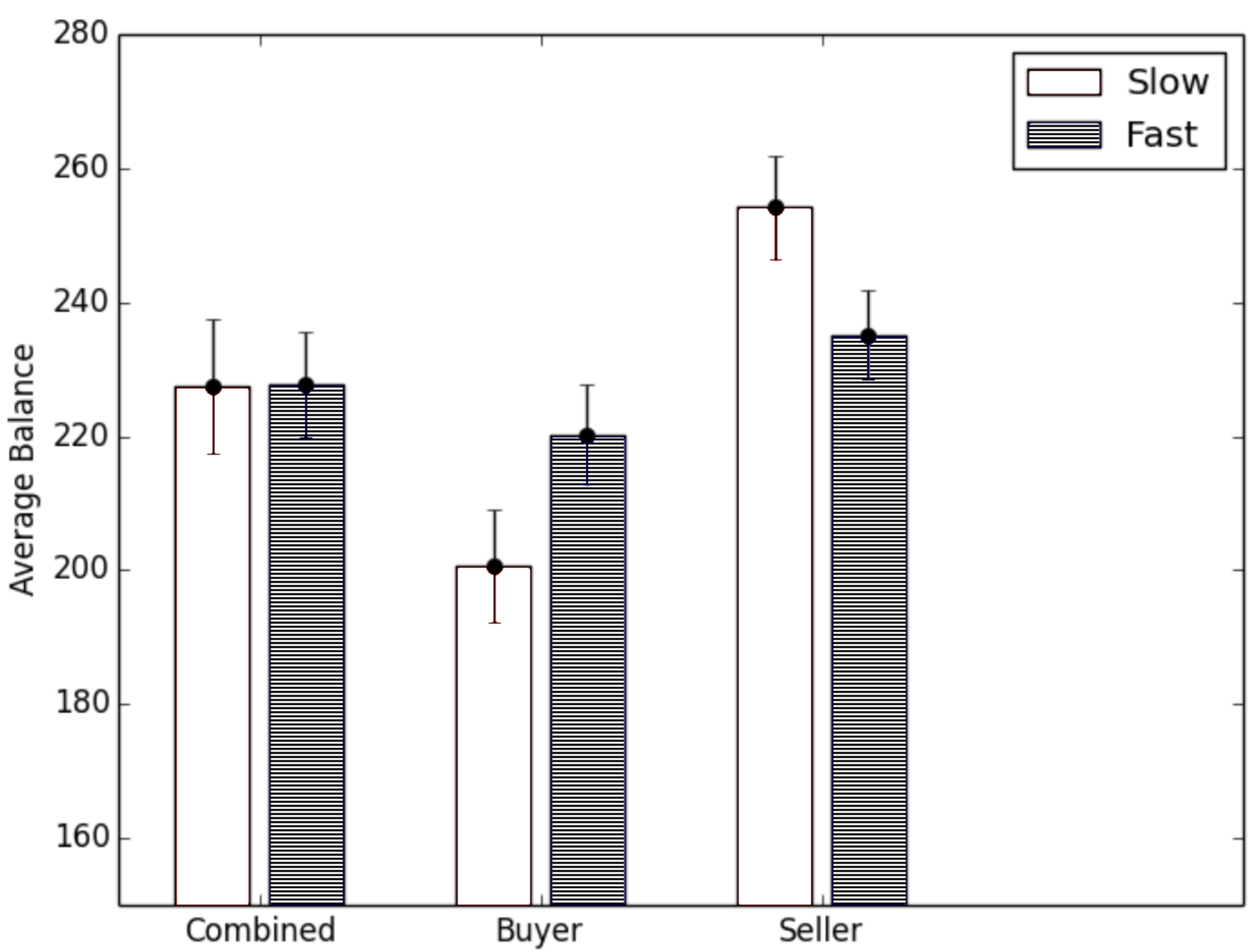}}\hspace{0.2\textwidth}
      \subcaptionbox{AA Quote Prices}[.32\linewidth][c]{%
        \includegraphics[width=\linewidth]{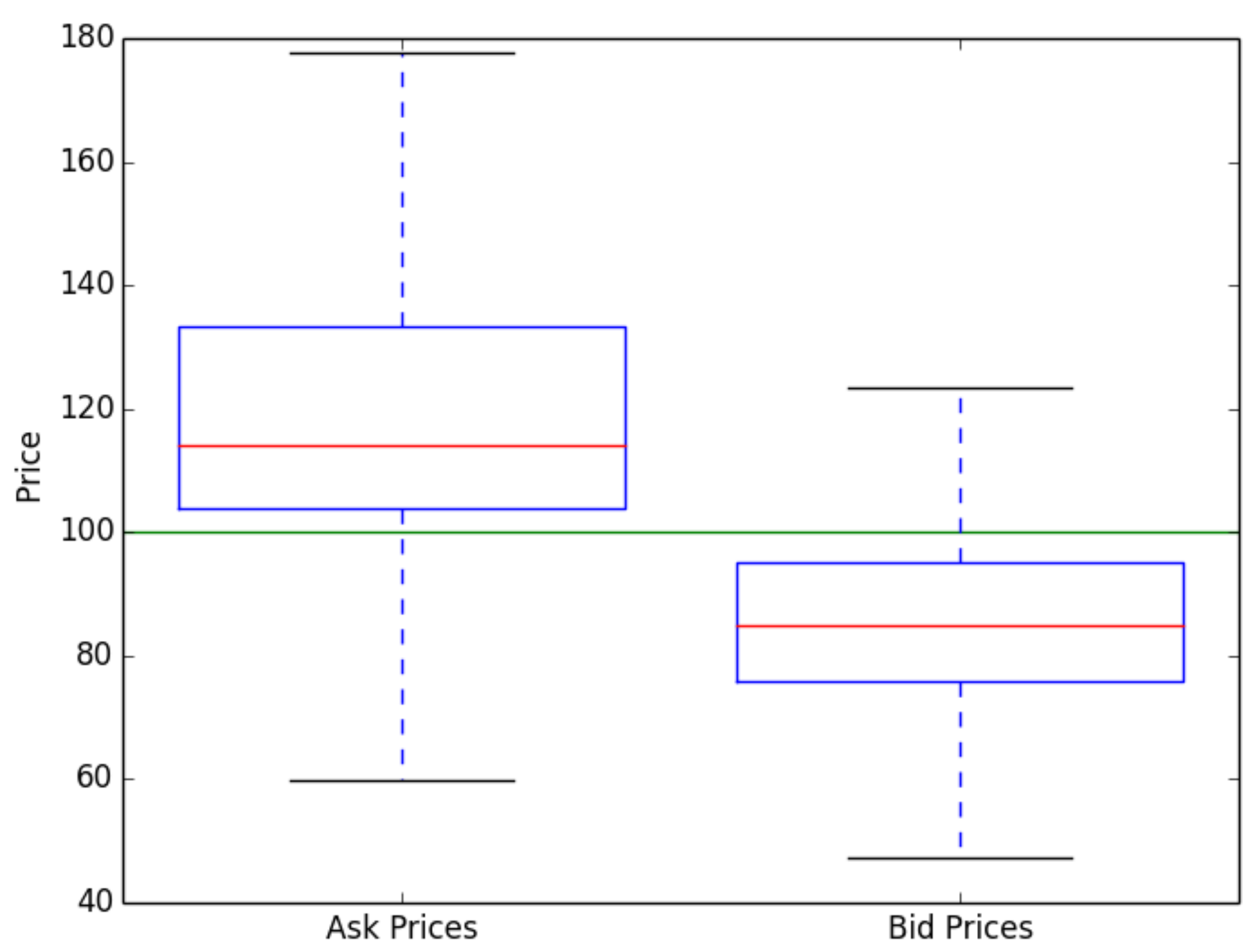}}
      \par\smallskip  
      \caption{Tournament ranking selection for homogeneous AA markets: (a) fast (grey) buyers outperform slow (white) buyers, while slow (white) sellers outperform fast (grey) sellers; (b) distribution of quote prices, showing asks (left box-plot) tend to be posted farther from equilibrium (green line) than bids (right box-plot). 
      } 
      \label{fig:AA-rank}
    \end{figure*}

\else
\begin{figure*}[t!]
    \centering
    \begin{subfigure}[t]{0.5\textwidth}
        \centering
         \includegraphics[width=0.65\linewidth]{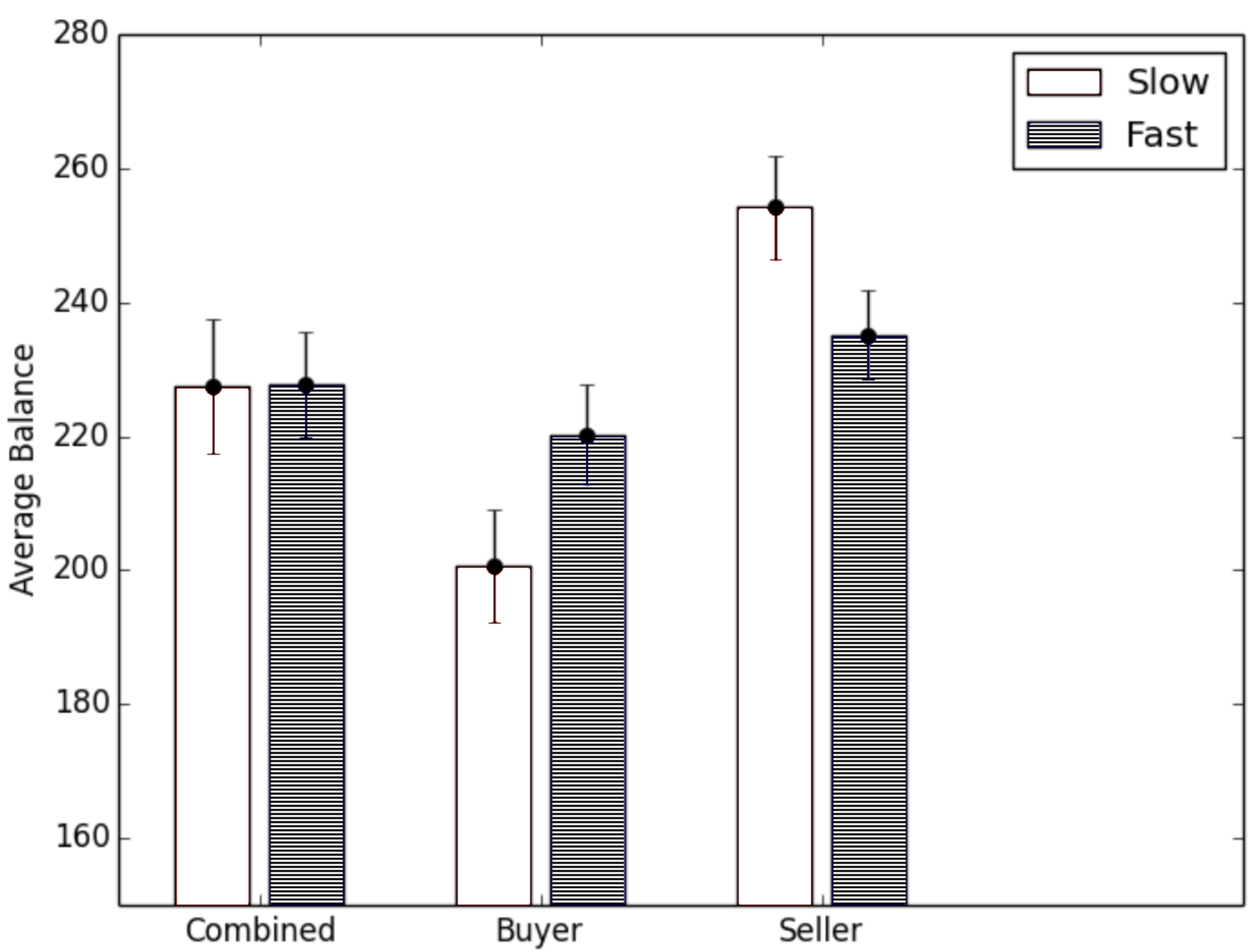}
        \caption{AA Profits}
    \end{subfigure}%
    ~ 
    \begin{subfigure}[t]{0.5\textwidth}
        \centering
        \includegraphics[width=0.65\linewidth]{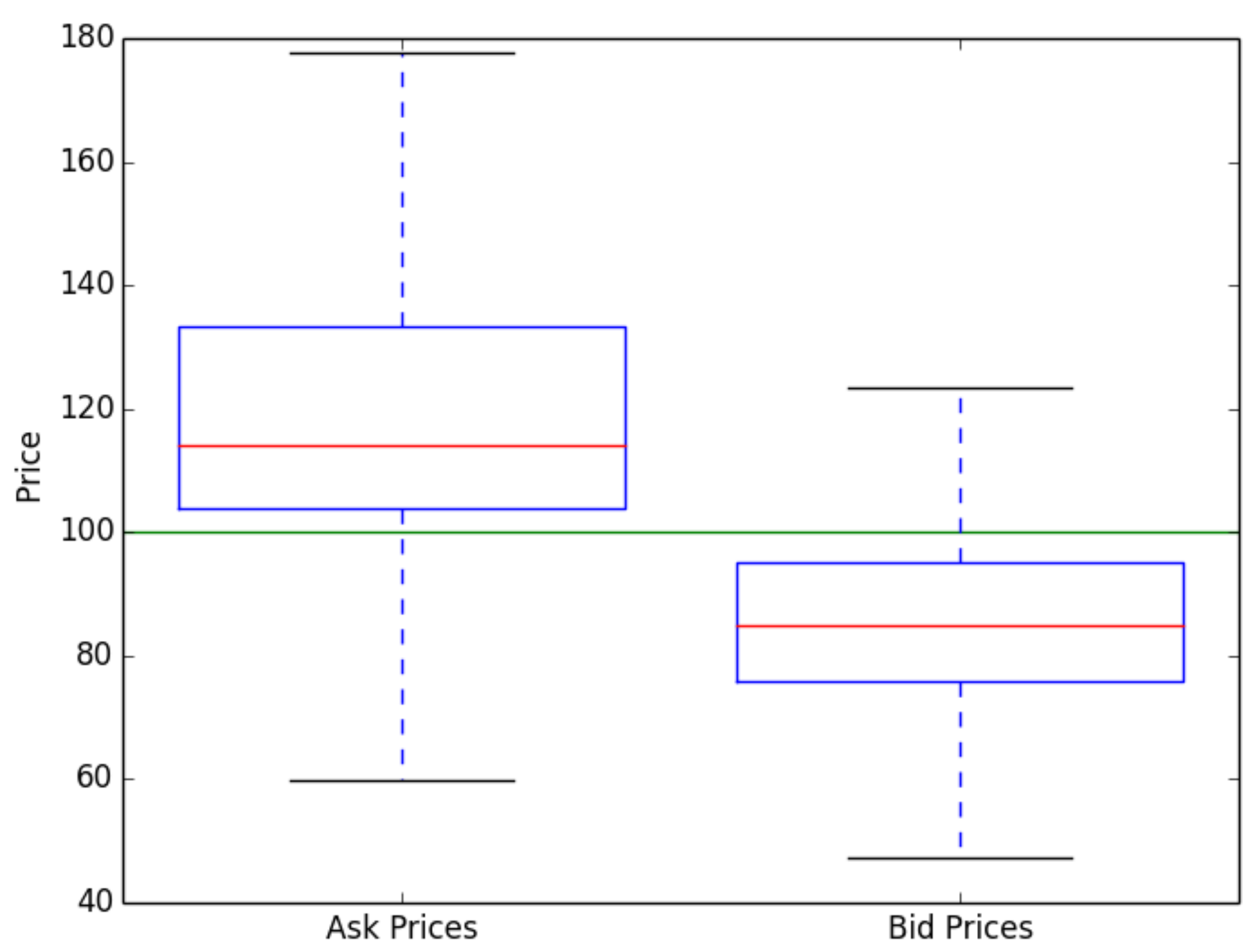}
        \caption{AA Quote Prices}
    \end{subfigure}
    \par\smallskip 
    \caption{Tournament ranking selection for homogeneous AA markets: (a) fast (grey) buyers outperform slow (white) buyers, while slow (white) sellers outperform fast (grey) sellers; (b) distribution of quote prices, showing asks (left box-plot) tend to be posted farther from equilibrium (green line) than bids (right box-plot).}
    \label{fig:AA-rank}
    \end{figure*}
\fi

\subsection{Fixed Order Results}
\label{sec:results-order}

We ran a series of experiments using homogenous trading populations with fixed order selection. Results demonstrate that the order in which traders act had no effect on traders' performance (measured by profit generated) for GVWY, SHVR, and ZIC markets (results not shown). This result is perhaps predictable, as these simple trading agents cannot identify and optimally capitalise on profit making opportunities, unless they do so accidentally.
However, a difference can be observed in results for homogeneous ZIP markets and homogeneous AA markets. 
In ZIP markets (Figure~\ref{fig:homo-order-ZIP}), there is a significant difference in performance between traders selected first (trader 0) and traders selected last (trader 9). However, perhaps surprisingly, while it is advantageous to be selected earlier when a buyer, it is better to be selected later when a seller. In AA markets (Figure~\ref{fig:homo-order-AA}), the ordering has less impact, but there is a general trend that being selected later improves performance. In particular, traders selected last generate significantly more profit than traders selected first. This could be because AA traders that act later each time step have additional information available to produce a better estimation of $P_0$, and therefore are likely to post a more profitable quote. However, this simple ordered model is very contrived and so we only present these results as evidence that selection ordering impacts profitability of the {\em adaptive} traders, AA and ZIP. We investigate further using more realistic models in the following sections.

\subsection{Tournament Ranking Results}
\label{sec:rank}

\subsubsection{Homogeneous Markets}
\label{sec:rank-homo}

We performed tournament ranking selection in homogeneous markets. When pitted against only traders of the same type, GVWY and ZIP perform better when faster, but the difference is not significant. For other traders there is no difference between fast and slow traders. However, for AA (see Figure~\ref{fig:AA-rank}), we see that fast buyers significantly outperform slow buyers, while slow sellers significantly outperform fast sellers. Overall, sellers generate significantly more profit than buyers, which is a result of prices tending to approach equilibrium from above. We see that asks (i.e., sell quotes) tend to be farther from equilibrium than bids (i.e., buy quotes), suggesting that sellers are retaining higher profit margins than buyers. Therefore, as there are fewer opportunities for buyers to get a good deal, being faster is beneficial. Conversely, for sellers, it is better to be slower, and wait for more attractive bids to arrive each time step before acting. 

\begin{figure}[tb]
  \centering
   \includegraphics[width=0.7\linewidth]{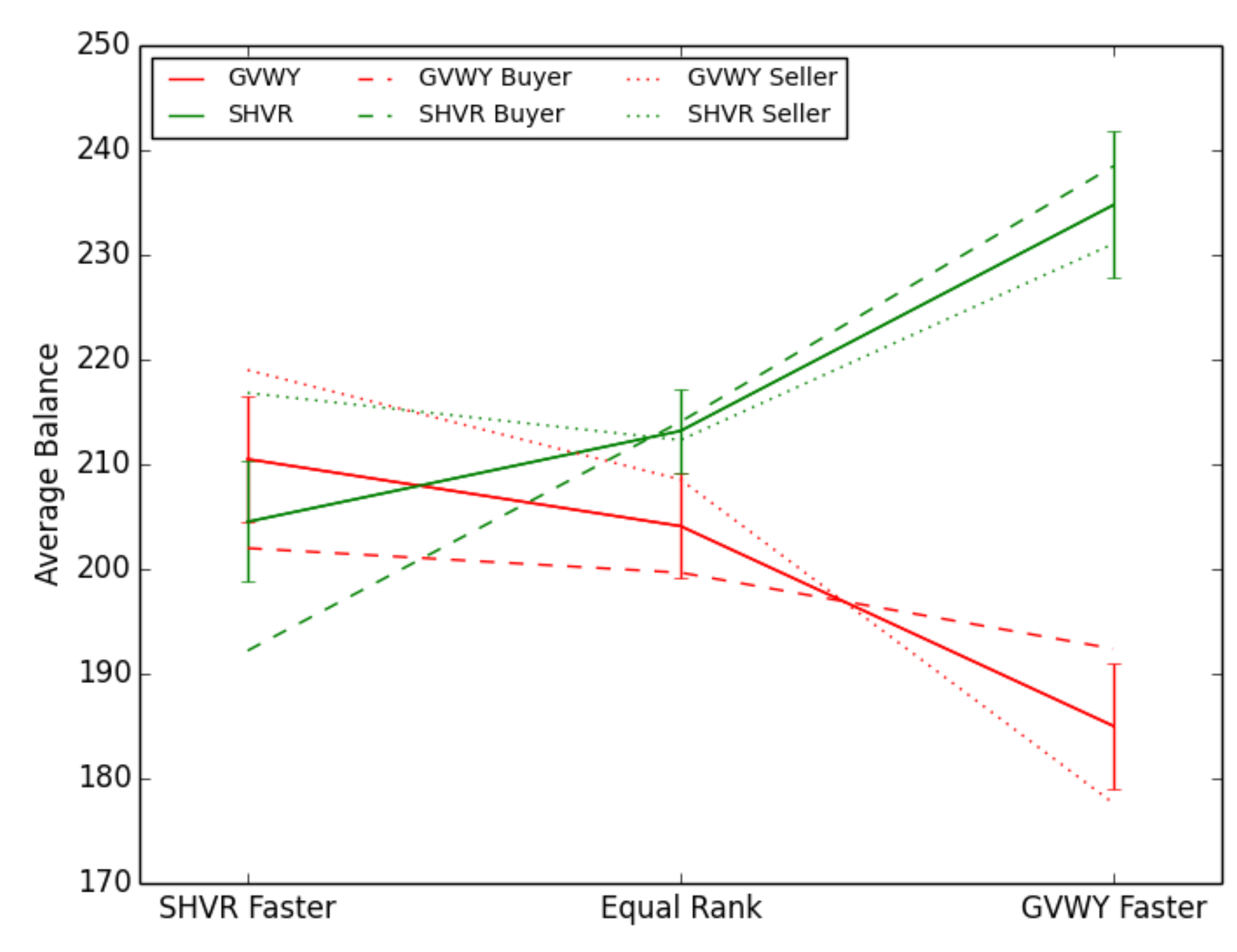}
  \caption{Balanced test SHVR:GVWY, using tournament ranking selection. 
  As GVWY (red) increases relative speed (left to right), SHVR (green) increasingly outperforms GVWY. 
  Both traders benefit from being ranked slowest.}
  \label{fig:SHVR-GVWY}
\end{figure}

\begin{figure}[tb]
  \centering 
   \includegraphics[width=0.68\linewidth]{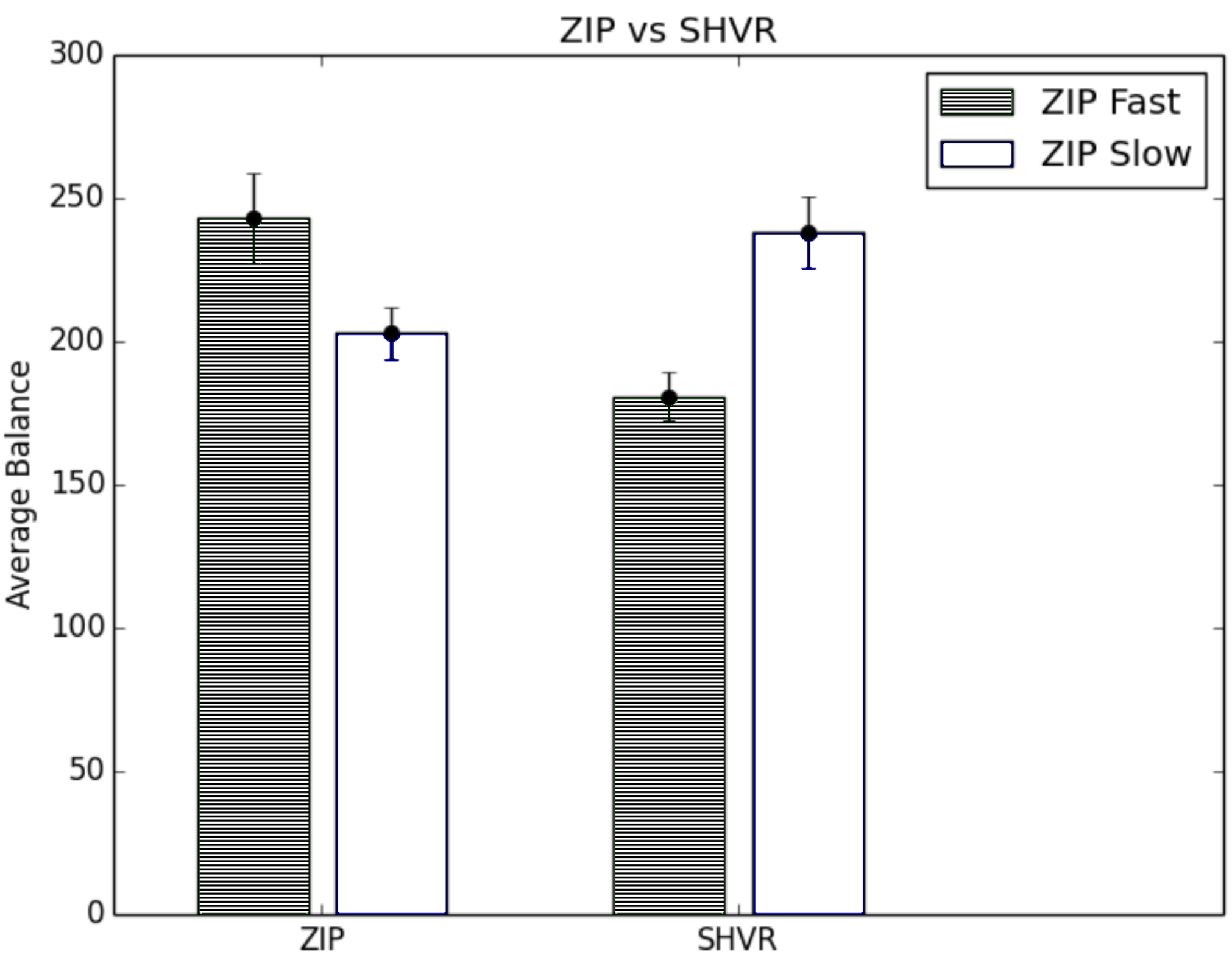}  
  \caption{Balanced test ZIP:SHVR, using tournament ranking selection. 
  ZIP significantly outperforms SHVR when ZIP is faster (grey bars), 
  but is significantly outperformed by SHVR when ZIP is slower (white bars).}
  \label{fig:hetero-rank-ZIP-SHVR}
\end{figure}

\subsubsection{Heterogeneous: Balanced Tests}
\label{sec:rank-hetero}

\ifnum\arXiv=0
\begin{figure*}[tb]
  \centering 
  \subcaptionbox{AA:GVWY}[0.35\linewidth][c]{%
    \includegraphics[width=\linewidth]{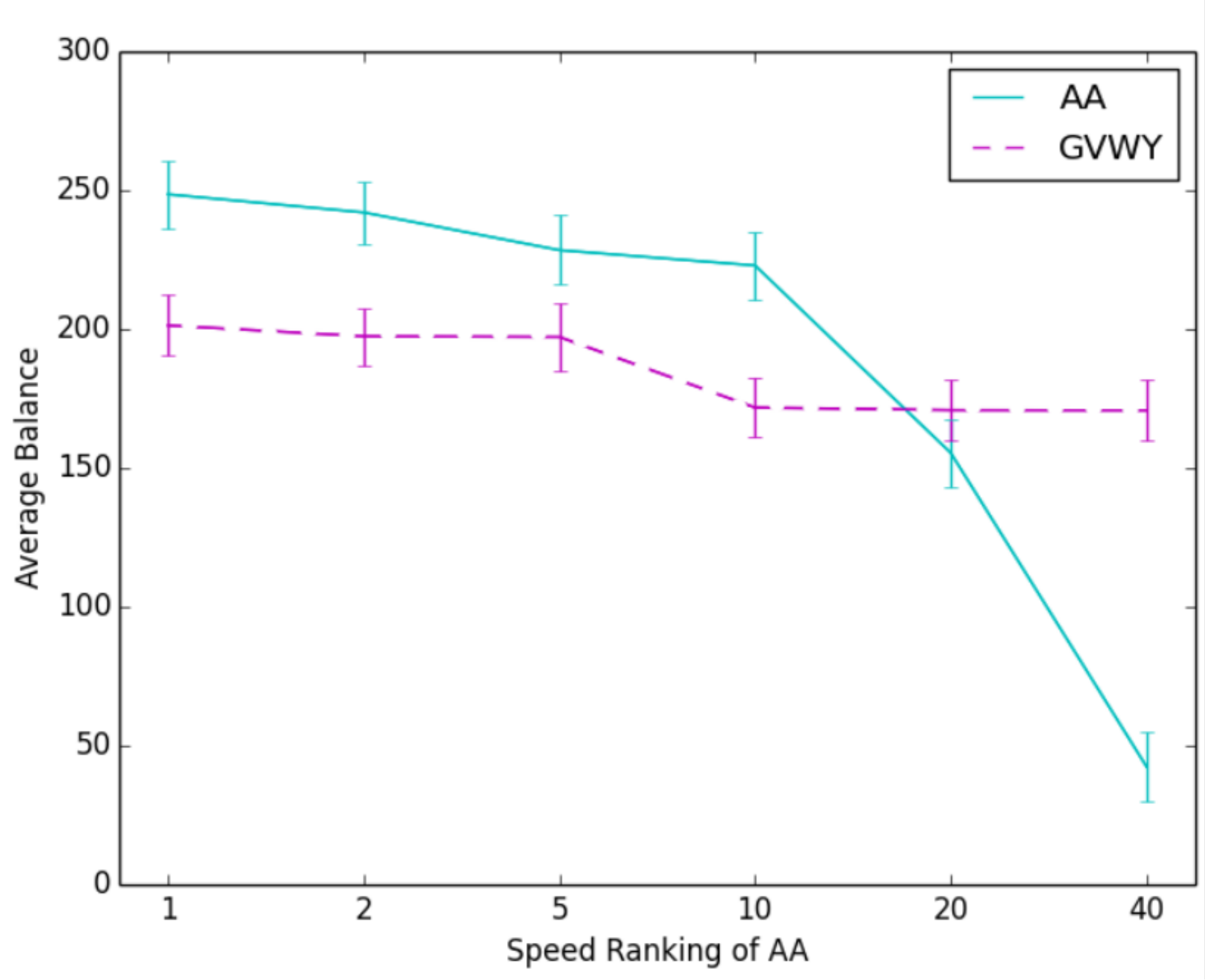}} \hspace{0.1\textwidth}
  \subcaptionbox{AA:SHVR}[.35\linewidth][c]{%
    \includegraphics[width=\linewidth]{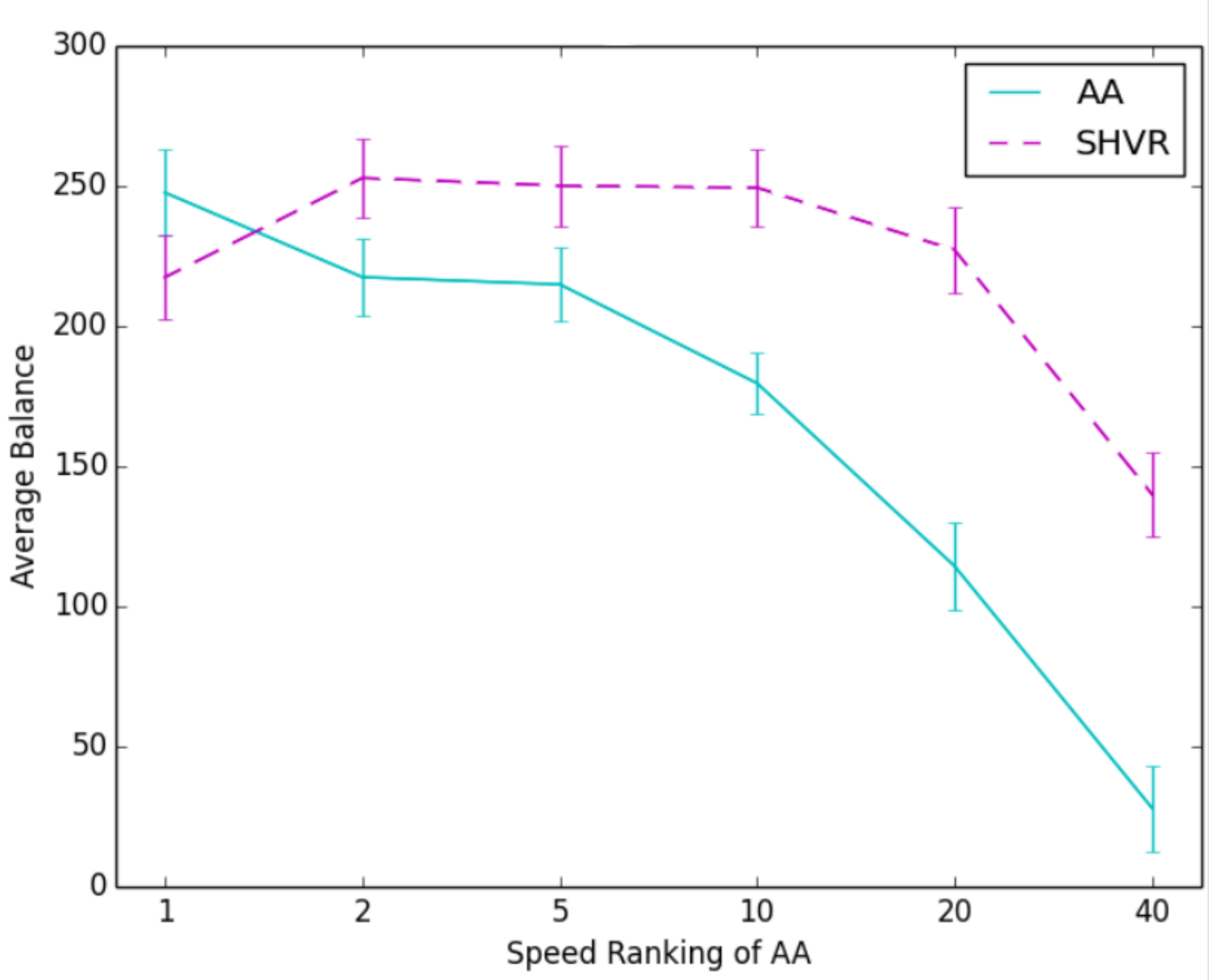}}
  \par\smallskip\par\smallskip
  \subcaptionbox{AA:ZIC}[.35\linewidth][c]{%
    \includegraphics[width=\linewidth]{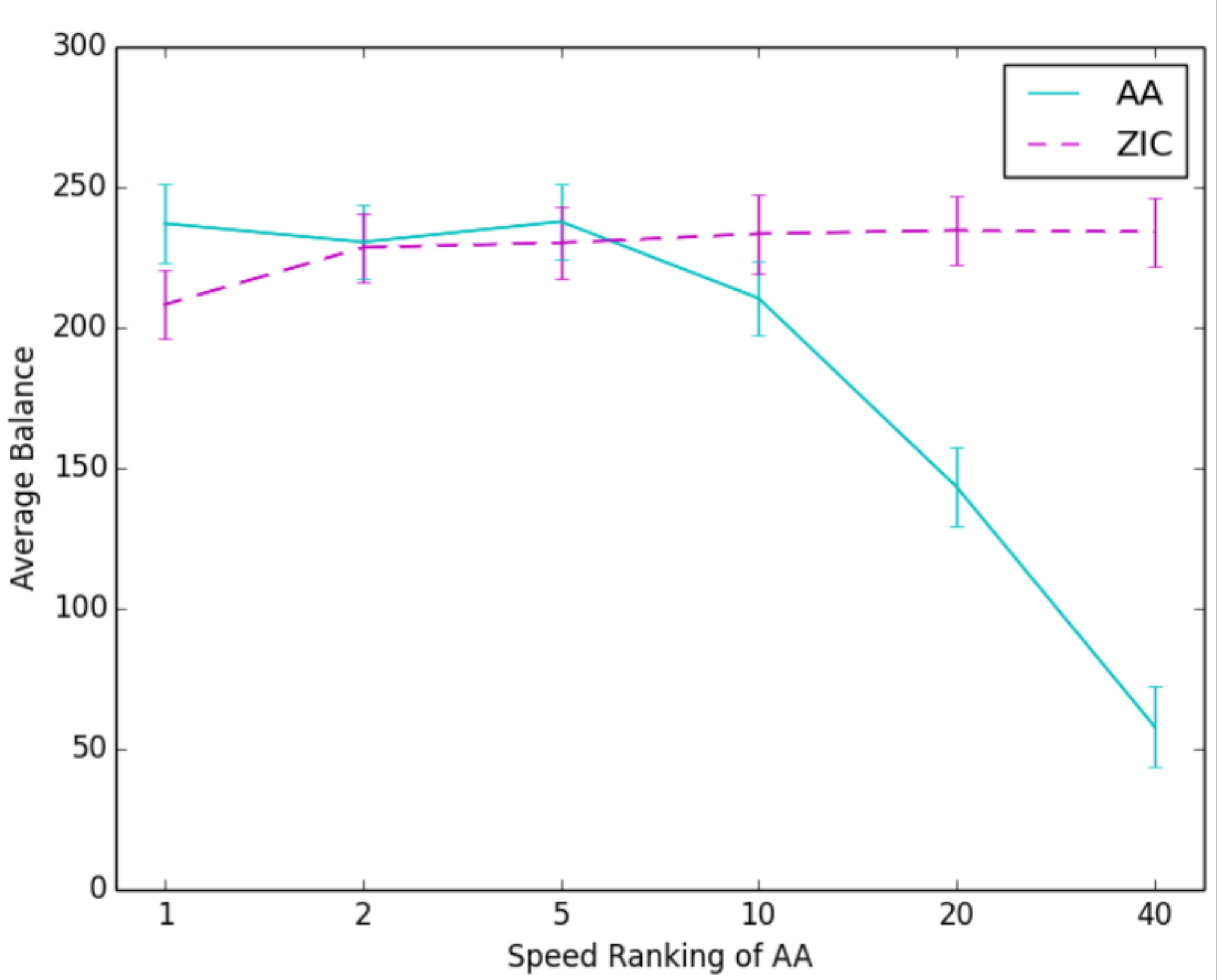}} \hspace{0.1\textwidth}
  \subcaptionbox{AA:ZIP}[.35\linewidth][c]{%
    \includegraphics[width=\linewidth]{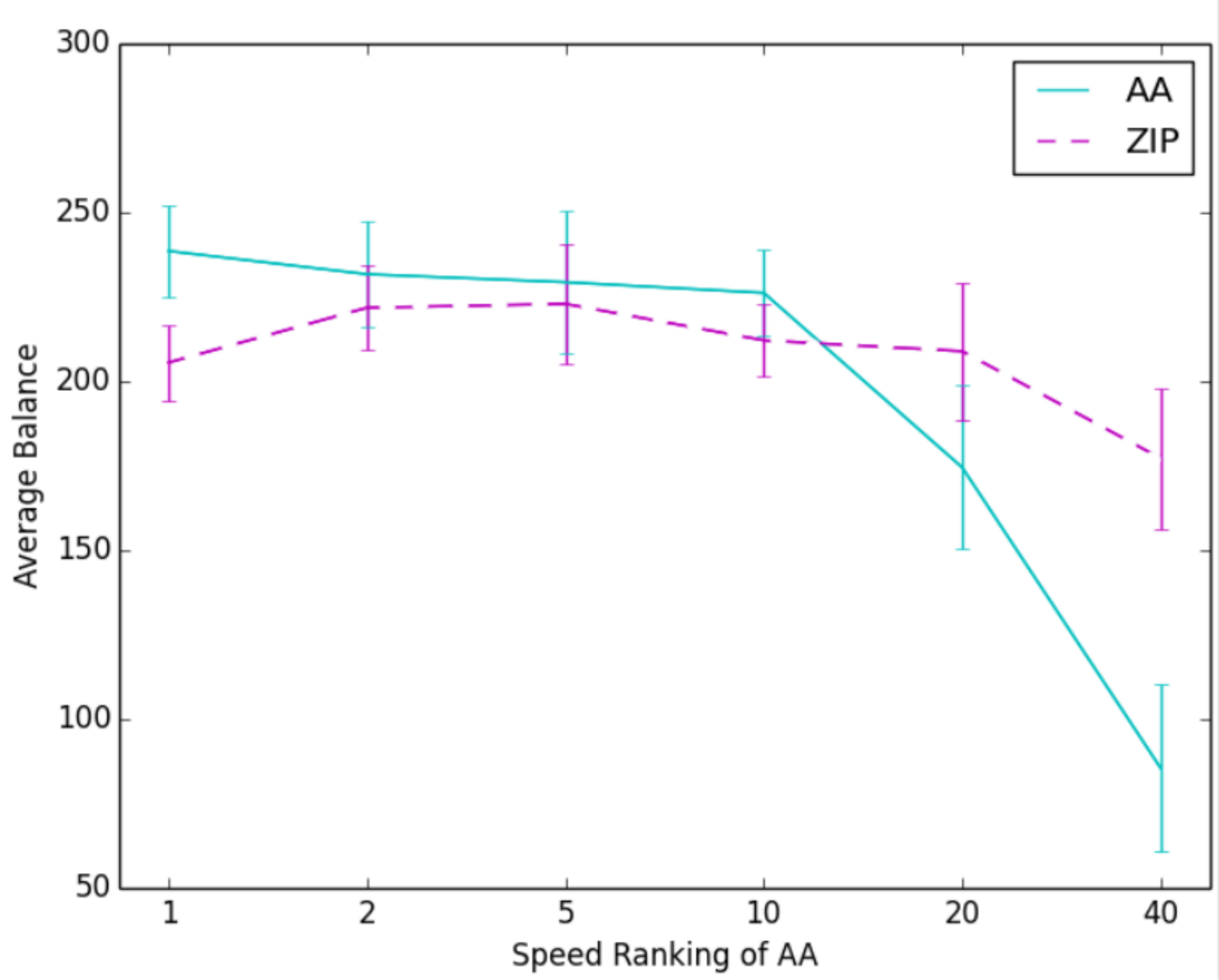}}
  \par\smallskip  
  \caption{Sensitivity analysis of AA using speed proportional selection in heterogeneous markets. 
  Reaction time of AA relative to the competing trader type is varied from $R^{AA}_*=1$ to $R^{AA}_*=40$ (x-axis). 
  Each test, AA (light blue) outperforms the competitor (purple dash) when compute times are equal ($R^{AA}_*=1$). 
  As $R_*^{AA}$ increases, 
  AA performance falls, until an inversion point is reached where AA no longer outperforms the competitor. 
  For SHVR, inversion occurs between $1<R_{SHVR}^{AA}<2$.}
  \label{fig:AA-timing}
\end{figure*}
\else
\begin{figure*}[t!]
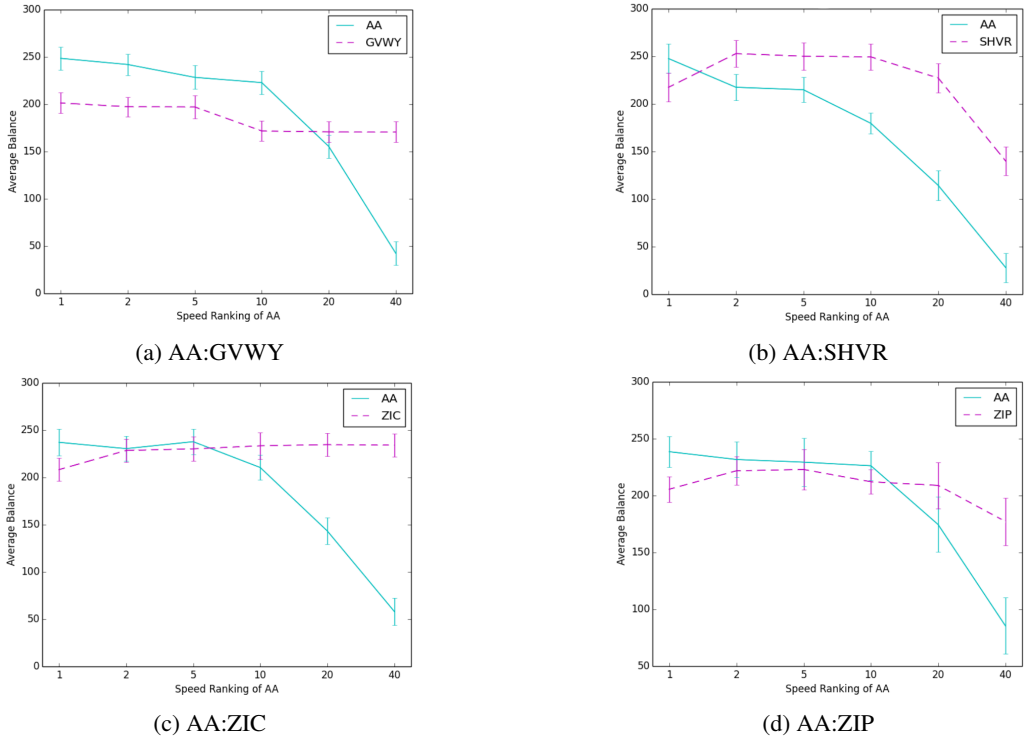

    \centering
    \begin{subfigure}[t]{0.49\textwidth}
        \centering
         \includegraphics[width=0.7\linewidth]{fig/AA-GVWY-alltimes-d}
        \caption{AA:GVWY}
    \end{subfigure}
    ~ 
    \begin{subfigure}[t]{0.49\textwidth}
        \centering
        \includegraphics[width=0.7\linewidth]{fig/AA-SHVR-alltimes-d}
        \caption{AA:SHVR}
    \end{subfigure}
    \par\smallskip
    \begin{subfigure}[t]{0.49\textwidth}
        \centering
        \includegraphics[width=0.7\linewidth]{fig/AA-ZIC-alltimes-d}
        \caption{AA:ZIC}
    \end{subfigure}
     ~
    \begin{subfigure}[t]{0.49\textwidth}
        \centering
        \includegraphics[width=0.7\linewidth]{fig/AA-ZIP-alltimes-d}
        \caption{AA:ZIP}
    \end{subfigure}       
    \par\smallskip
  \caption{Sensitivity analysis of AA using speed proportional selection in heterogeneous markets. 
  Reaction time of AA relative to the competing trader type is varied from $R^{AA}_*=1$ to $R^{AA}_*=40$ (x-axis). 
  Each test, AA (light blue) outperforms the competitor (purple dash) when compute times are equal ($R^{AA}_*=1$). 
  As $R_*^{AA}$ increases, 
  AA performance falls, until an inversion point is reached where AA no longer outperforms the competitor. 
  For SHVR, inversion occurs between $1<R_{SHVR}^{AA}<2$.}
  \label{fig:AA-timing}
    \end{figure*}
\fi

In many balanced tests there was no significant differences in outcomes between trader strategies when varying speed using the ranking model (not shown).  
However, results for SHVR are interesting. For GVWY:SHVR (see Figure~\ref{fig:SHVR-GVWY}), it can be seen that it benefits SHVR to be {\em slower} than GVWY. 
GVWY is an {\em honest} strategy (always quoting at the current limit price) with no consideration of the order book. Therefore, SHVR benefit from na\"ive GVWY traders posting earlier each time step. 
For ZIP:SHVR (see Figure~\ref{fig:hetero-rank-ZIP-SHVR}), we see that faster ZIP outperform slower SHVR, while faster SHVR outperform slower ZIP.  This is the result that we may intuitively expect, if we consider that faster is always better. 

These results indicate that the benefits of speed depend on the competing strategies in the market. For the parasitic SHVR, which requires a `sensible' order book to trade sensibly, it is better to be fast when intelligent traders are quoting in the market. Conversely, when competitors are trading honestly, it is beneficial to be slower. 
However, while the ranking model enables traders to directly compete for opportunities on speed alone, it does not allow faster traders to act more often than slower traders, which is unrealistic. We remove this constraint in the following model. 

\subsection{Speed Proportional Results}
\label{sec:time}

\subsubsection{AA: Reaction Speed Sensitivity Analysis}

Figure~\ref{fig:AA-timing} presents results of speed sensitivity analysis of AA performance in heterogeneous balanced-group tests against each of the other four trading strategies: (a) GVWY, (b) SHVR, (c) ZIC, (d) ZIP. The reaction time of AA relative to the competing trader type is varied from $R^{AA}_*=1$ to $R^{AA}_*=40$. Graphs show the effect of increasing $R^{AA}_*$ (x-axis). In each case, we see that when $R^{AA}_*=1$ (i.e., equal reaction times), AA (light blue line) outperforms the competing trader (purple dashed line). This is the BSE default setting, and the result conforms to previous findings that $AA$ dominates in symmetric markets with balanced numbers of traders \citep{DeLucaCliff11}.

However, as relative AA reaction time $R^{AA}_*$ is increased, we see that AA performance gradually falls, until a point is reached where $AA$ is beaten by the competing trader group. This inversion point varies between trader groups, but occurs very quickly for SHVR, between $R^{AA}_{SHVR}=1$ and $R^{AA}_{SHVR}=2$, suggesting that AA's dominance over SHVR is sensitive to small variations in relative trader speeds.

\subsubsection{Profiling Reaction Times of Traders}
\label{sec:profiling} 
Here, we attempt to accurately profile the reaction times of each trading agent. In BSE, the computation time of an agent is composed of two methods: \texttt{getOrder}, which is called each time a trader is selected to submit a new order into the market, thus requiring the calculation of a new quote price, $Q$; and \texttt{respond}, which is called after each market event, and is used by traders to update internal parameters based on the event data (e.g., a new trade, or a new best bid or ask on the LOB).

Of the trading agents considered in this work, three are stateless: GVWY, SHVR, and ZIC. These traders have no internal parameters to update and therefore take no action when their \texttt{respond} method is called. Only ZIP and AA have an internal state. These traders use their \texttt{respond} method to update internal variables in order to calculate a new profit margin, $\mu$. When \texttt{getOrder} is called, ZIP and AA use their current profit margin, $\mu$, to calculate a new quote price, $Q=\mu L$. In comparison, GVWY and ZIC generate a quote price without reference to market data; SHVR uses the current best bid and ask in the LOB to generate a new quote price. 

Table~\ref{tab:profile} shows the profiled reaction times of each trading agent, observed across 52 million method calls under varying market conditions including population size, mix of traders in the market, assignment replenishment schedules, etc. 
\ifnum\ANON=0 
(see \cite{Hanifan19} for details). 
\else
(see {\em redacted} for details).
\fi
Unsurprisingly, we see that the traders with the longest computation times, ZIP and AA, are those with an internal state that requires continuous updating. 
The relative reaction times between the fastest and slowest traders is roughly a factor of two: $R_{GWVY}^{AA}=2.26$. 
The relative reaction times between AA and ZIP is $R^{AA}_{ZIP}=1.13$. This result is consistent with the average relative times $R^{AA}_{ZIP}=1.19$ presented by \citet{SnashallCliff20}, and we take this as confirmatory evidence that our profiling is accurate. The final column of Table~\ref{tab:profile}, headed $R^*_{SHVR}$, presents reaction time of each trader relative to SHVR. Generating a quote price relative to the current LOB, SHVR is the only stateless (and therefore {\em fast}) trader that reacts to market information; although it does so in a simplistic non-adaptive fashion (unlike the slower AA and ZIP). 

\begin{table}[tb]
\vspace{2.5mm}
\caption{Profiled reaction times of trading agents.}\label{tab:profile} 
  \small
  \centering
  \begin{tabular}{ccccc}
  \toprule
  Trader  & Time ($\mu$s) & Stateful & Reactive & $R^*_{SHVR}$\\
  \midrule
  GVWY & 4.2 & N & N & 0.61\\
  SHVR  & 6.9 & N & Y & 1.00\\ 
  ZIC      & 7.1 & N & N & 1.03\\
  ZIP      & 8.4 & Y & Y & 1.22\\
  AA       & 9.5 & Y & Y & 1.38\\
  \bottomrule
\end{tabular}
\end{table}

\ifnum\arXiv=0
\begin{figure*}[tb]
  \centering 
  \label{subfig:a}
  \subcaptionbox{ZIP:SHVR}[0.4\linewidth][c]{%
    \includegraphics[width=\linewidth]{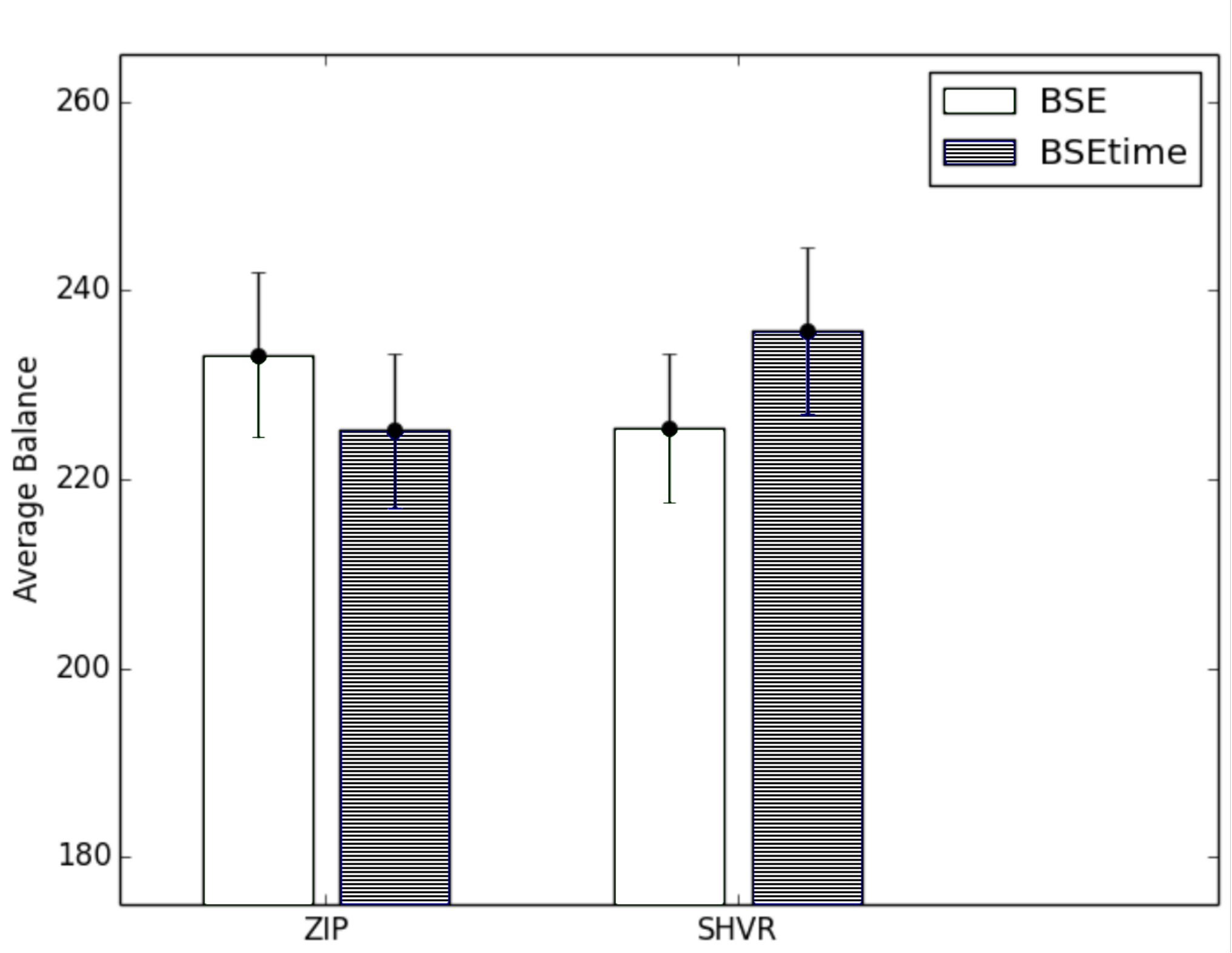}} \hspace{0.1\textwidth}
  \subcaptionbox{AA:SHVR}[0.4\linewidth][c]{%
    \includegraphics[width=\linewidth]{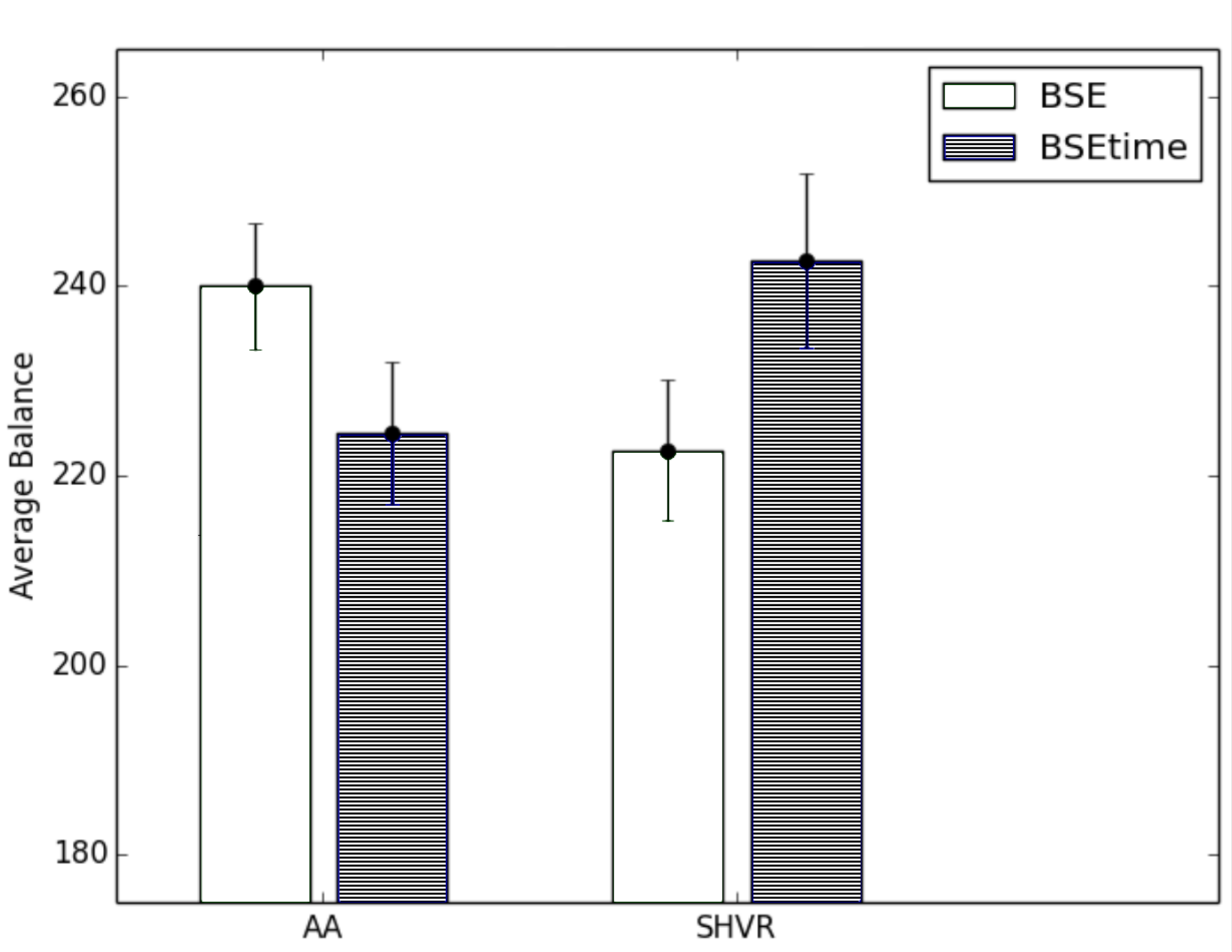}}
  \par\smallskip
  \caption{Comparison of results between random selection (white) and speed proportional selection (grey), for heterogeneous markets using profiled thinking times (see Table~\ref{tab:profile}): (a) SHVR outperforms ZIP under speed proportional selection (not significant); (b) SHVR significantly outperforms AA under speed proportional selection.}
  \label{fig:hetero-time}
\end{figure*}
\else
\begin{figure*}[t!]
    \centering
    \begin{subfigure}[t]{0.49\textwidth}
        \centering
         \includegraphics[width=0.8\linewidth]{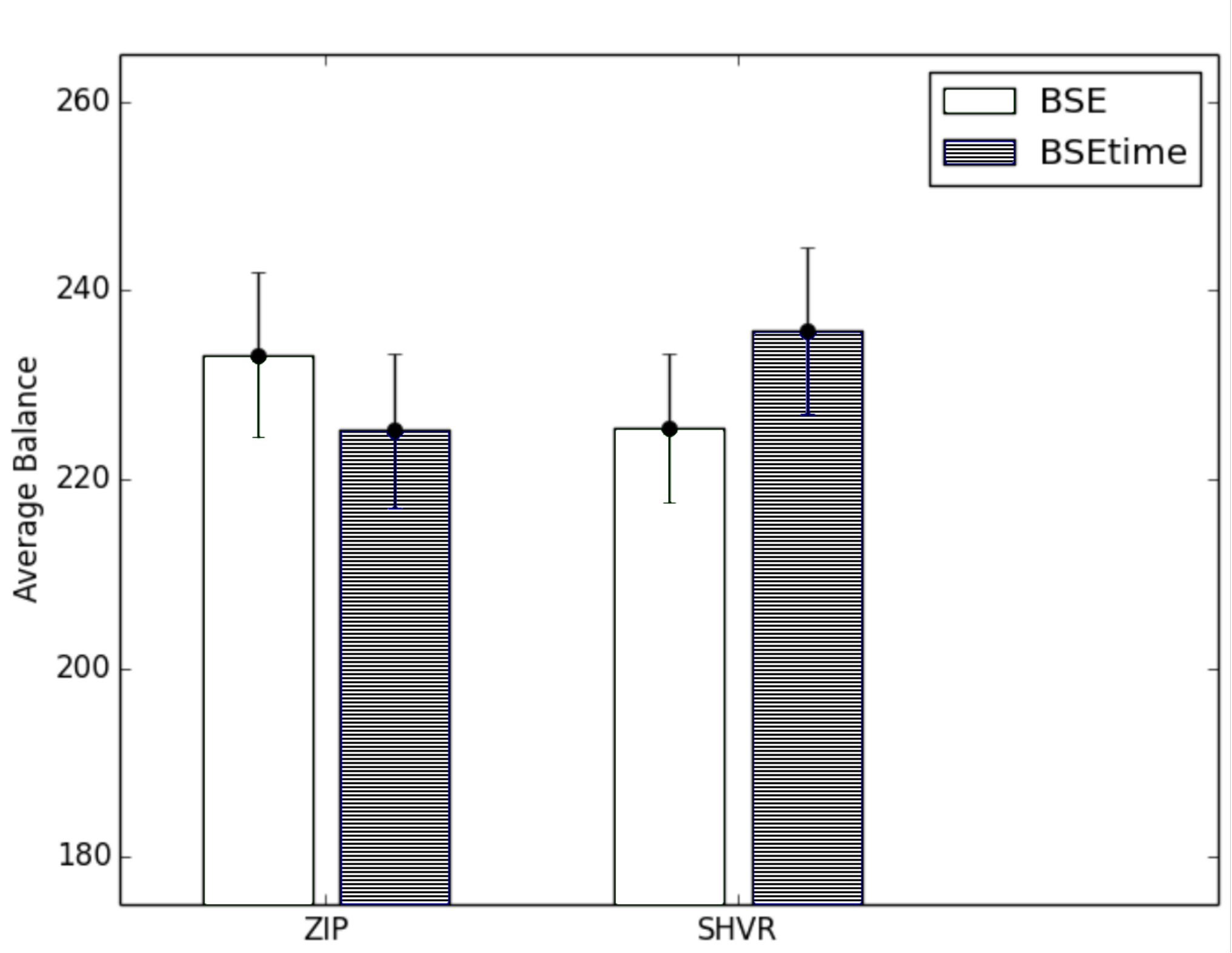}
        \caption{ZIP:SHVR}
    \end{subfigure}
    ~ 
    \begin{subfigure}[t]{0.49\textwidth}
        \centering
        \includegraphics[width=0.8\linewidth]{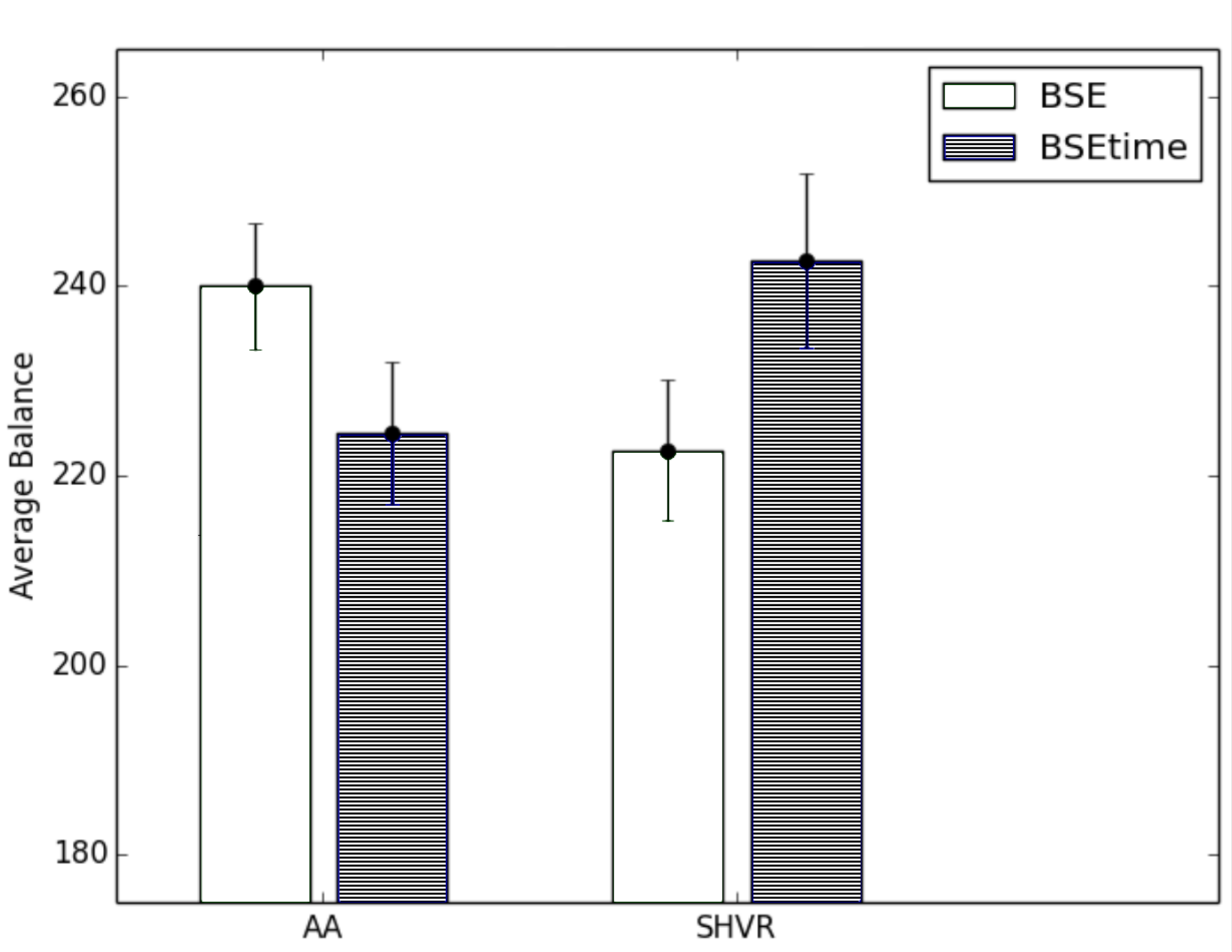}
        \caption{AA:SHVR}
    \end{subfigure}      
    \par\smallskip
  \caption{Comparison of results between random selection (white) and speed proportional selection (grey), for heterogeneous markets using profiled thinking times (see Table~\ref{tab:profile}): (a) SHVR outperforms ZIP under speed proportional selection (not significant); (b) SHVR significantly outperforms AA under speed proportional selection.}
  \label{fig:hetero-time}
\end{figure*}
\fi

\subsubsection{Results Using Profiled Reaction Times}
We used profiled computation times (Table~\ref{tab:profile}) for proportional selection in heterogeneous balanced-group tests for pairwise comparisons between all trader types. The majority of results showed no significant difference, suggesting the relative differences in reaction speeds between the trader agents are not large enough to have an impact. However, results for ZIP:SHVR and AA:SHVR were particularly interesting (see Figure~\ref{fig:hetero-time}). 
For ZIP:SHVR (Figure~\ref{fig:hetero-time}(a)), under BSE's default randomised selection process (white bars), ZIP outperforms SHVR. However, when selecting traders proportional to their true relative speeds (grey bars) SHVR outperforms ZIP (although the difference is not significant). A similar, but more pronounced trend emerges between AA:SHVR (Figure~\ref{fig:hetero-time}(b)). Here, AA significantly outperforms SHVR under the default randomised selection (white), and significantly underperforms SHVR under speed proportional selection (grey). 

\section{\uppercase{Discussion}}
\label{sec:discussion}
\noindent 
The result presented in Figure~\ref{fig:hetero-time}, demonstrating SHVR is more profitable than AA (significantly, $p<0.05$) and ZIP (not significant, $p>0.05$) is a novel result. By accurately accounting for the relative reaction times of the two algorithms, we have demonstrated that, in balanced tests, 
SHVR---the simple non-adaptive order book strategy---is able to generate more profit than AA in a Smith-style static symmetric marketplace, of the kind that AA was specifically designed to succeed in \citep{Vytelingum06}, and in which several studies have previously demonstrated AA as being the dominant known strategy \citep{Vytelingum08,DeLucaCliff11}. 

More recently, the dominance of AA has been questioned in several works. Using the discrete-event simulation mode of OpEx, \cite{Vach15} used Smith-style markets (similar to those used here) to compare efficiencies of traders in markets containing AA, GDX, and ZIP, as the proportion of each trader type in the market was varied. He showed that for large regions of the mixture space, GDX was the dominant strategy in these 3-way markets. Later, using BSE, \cite{Cliff19} demonstrated that in Smith-style markets containing equal proportions of MAA (AA modified to use {\em microprice}), SHVR, ZIC, and ZIP, profits per trader have no significant difference; while in more complex markets with continuously varying equilibria, SHVR and ZIP significantly outperform MAA on profits. A follow-up study \citep{SnashallCliff20} demonstrated that GDX dominates in complex markets containing MAA, ASAD, GDX, and ZIP; and on average scores greater efficiency than AA in the simpler Smith-style markets. These works show that AA's previously perceived dominance is sensitive to the mixture of competing strategies in the market, and the complexity of market dynamics. Here, we were able to demonstrate that even in Smith-style markets with two balanced groups of traders (the exact markets that AA was previously shown to dominate), AA is less profitable than the simple SHVR strategy when we account for reaction speed. 

We believe that this finding is significant, not only because it contributes to the recent body of evidence suggesting that AA is non-dominant, but also because it demonstrates that the performance of adaptive trading algorithms (AA and ZIP) are sensitive to reaction time; and once reaction time is considered, SHVR may be relatively superior. In more complex markets designed to emulate real-world financial dynamics, SHVR has previously been shown to outperform AA and to perform similarly to ZIP \citep{Cliff19}. Here, we extend this result to show that SHVR can also outperform in simple markets, once we account for speed. In this study, we have not considered the GDX trading strategy and we reserve this for future work. However, we note that \cite{SnashallCliff20} have recently profiled the reaction time of GDX and shown it to be an order of magnitude slower than AA. We therefore believe that, given the evidence we have presented here, if we factor in speed using proportional selection, GDX would likely perform less well than SHVR (and also AA and ZIP), as each competing trader strategy would be able to act ten times for every GDX trader's action.

All evidence is starting to suggest that SHVR, although extremely simple, could be a profitable trading strategy to employ. However, we should perhaps not be too surprised. SHVR is the only agent described in this paper that directly uses the current order book to determine quote price (by simply shaving one tick off the current best bid or ask).  While AA, GDX, and ZIP are able to trade in markets containing an order book, metrics such as the current spread, volumes, buy/sell imbalances, etc., are not considered. This is valuable information that is routinely used by HFT strategies in the real markets. We therefore suggest that more emphasis is placed on utilising order book metrics for trading algorithms. We note that the MAA ({\em modified} AA) strategy introduced by \cite{Cliff19} considers the order book {\em microprice} to perform calculations for updating profit margin. This is a useful start, however we believe that this approach should be taken further, such that strategies should separate into two processes the long-term function of determining a desired price (e.g., by updating profit margin), and the short-term function of working to achieve that price (e.g., by shaving the bid). One could easily imagine a simple combination of ZIP and SHVR that could perform this strategy. We aim to investigate the incorporation of order book metrics into trading agent strategies in future work. 

In real-world markets, order book information and reaction speed is so strategically useful that, in order to stop a trader's trading {\em intention} from being used adversely by predatory competitors, some trading venues, described as {\em dark pools}, do not reveal quotes (see, e.g., \cite{MPCDark19}, for a summary of dark pools and methods for implementing cryptographically secure dark pool mechanisms using multi-party computation (MPC)). 
Further, as the majority of predatory HFT strategies rely on being quick(est) to act, there is also some movement in real markets towards introducing artificial delays ({\em speed bumps}), non-continuous trading, and re-ordering policies (see, e.g., \cite{Libra19} for a study on the effects of a {\em temporal fairness} policy on the {\em Refinitiv Matching} foreign exchange).

In real markets, communications latencies can be orders of magnitude larger than the reaction times of automated execution algorithms. However, these latencies can be (relatively easily) minimised by paying for (costly) services such as exchange hosting (enabling traders to co-locate equipment within the same data centre as the exchange matching engines and market data systems for the lowest latency access) and direct market access feeds. Therefore, for all automated trading systems (ATS) making use of these low-latency services (including the majority of HFT), communications latency becomes a level playing field. In these circumstances, reaction time is the principle speed differentiator. 

\section{\uppercase{Conclusions}}
\label{sec:conclusion}
\noindent 
We have introduced methods for simulating the reaction time of trading agents in financial market experiments, using the open-source {\em Bristol Stock Exchange} (BSE) platform. 
Historically, trading agent experiments have not considered the time it takes for agent strategies to compute: each time step, agents are selected to act in random order, and agents are unrestricted in the (real-world) time taken to compute an action. This is unrealistic, particularly given the unquestionable emphasis placed on trading speed in the real financial markets.

To simulate reaction time, three models were considered: (i) fixed ordering of agents; (ii) tournament selection using speed rankings; and (iii) proportional selection of agents relative to speed. The latter model---the most realistic---enabled fast traders to act multiple times for each individual action taken by a slower agent. We demonstrated, unsurprisingly, that speed does affect trader performance, although being faster is not always better: the outcome depends on the competing trader strategies present in the market; sometimes it can be beneficial to let the fools rush in.  

Significantly, we were able to show that when simulating accurate reaction times of agent strategies, the simplistic SHVR strategy is more profitable than AA. Until relatively recently, AA had been considered the dominant trading algorithm in the published literature, and this is the first time AA has been shown to be outperformed by a simple non-adaptive strategy in a static symmetric market containing two strategy types with equal numbers. 

This result confirms that speed matters, and if we are to better understand real-world markets it is necessary for reaction times to be considered more thoroughly. It also demonstrates that simple order book strategies can outperform intelligent trading strategies that make use of machine learning. This suggests that there should be further investigation into methods for incorporating order book information into adaptive trading strategies.

\section*{\uppercase{Acknowledgements}}
\noindent
\ifnum\ANON=0 
  John Cartlidge is sponsored by Refinitiv.
\else
  {\em Redacted for double-blind review.}
\fi

\atColsBreak{\vskip6pt} 

\bibliographystyle{apalike}
{\small
\bibliography{icaart}}

\end{document}